\documentclass[12pt]{iopart}

\usepackage{bbm}
\newsavebox{\myhbar}
\savebox{\myhbar}{$\hbar$}
\usepackage{tensor}
\usepackage{amssymb}
\usepackage{bm}
\usepackage[makeroom]{cancel}
\usepackage{xcolor}
\usepackage{cite}

\begin{document}
\title{Gravitational decoherence of photons}
\author { Michalis Lagouvardos and Charis Anastopoulos    \\
 {\small Department of Physics, University of Patras, 26500 Greece} }

\ead{mikelagou@upatras.gr, anastop@upatras.gr}
\begin{abstract}
Models of gravitational decoherence are not commonly applied to ultra-relativistic systems, including photons. As a result,  few quantum optical tests of gravitational decoherence have been developed. In this paper, we generalize the gravitational decoherence model of Anastopoulos and Hu \cite{AnHu13} to photons. In this model, decoherence originates from a bath of stochastic gravitational perturbations, possibly of fundamental origin. We derive a master equation for general states of the electromagnetic (EM) field; the only free parameter is a noise temperature $\Theta$ of the gravitational fluctuations. We find that interference experiments with long baselines, accessible in near-future experiments, can, in principle, lead to strong constraints in $\Theta$.
 
\end{abstract}

\section{Introduction}
Gravitational decoherence refers to the loss of coherence in matter due to gravity.   It has received strong attention in recent years, partly because of the possibility to test leading models, see, for example, Refs. \cite{MSPB, maqro, HSMC, PC20}.

Most gravitational decoherence models focus on the non-relativistic regime, i.e., they consider position or momentum decoherence for  particles at small velocities. There are few, if any, extensions of   popular decoherence models to ultra-relativistic systems. Hence, few quantum optical tests for gravitational decoherence have been considered. This problem originates from the fact that most gravitational decoherence models do not employ a Quantum Field Theory (QFT) description of matter, hence, they do not easily generalize to relativistic systems.
One exception is the gravitational decoherence model of Anastopoulos and Hu \cite{AnHu13} and of Blencowe \cite{Blencowe}, henceforward referred to as the ABH model.
 
 In this paper, we extend the ABH model to  the quantum electromagnetic (EM) field. We construct an effective master equation for the EM-field degrees of freedom in presence of a background of stochastic gravitational perturbations. The master equation depends on a single parameter, the strength of the stochastic perturbations. This parameter  can be plausibly constrained by quantum optics experiments in deep space.

\medskip


The original motivation for gravitational decoherence models is twofold \cite{Karol, Karol2, Penrose1, Diosi1,  Diosi2}. First, the existence of a fundamental decoherence process could be a solution to the quantum measurement problem, and  it could explain the emergence of a classical macroscopic world. Second, if the gravitational field is not fundamentally quantum, then gravity could interact with quantum systems through channels other than Hamiltonian evolution, possibly leading to decoherence of matter degrees of freedom. Hence, an experimental verification of gravitational decoherence could provide significant novel information about the fundamental nature of gravity.

The most influential model of gravitational decoherence is  the Diosi-Penrose (DP) model.  Penrose proposed that    spatial quantum superpositions collapse as a result of  back-reaction from gravity, due to the fact that the spacetime geometry is different in each branch of the superposition \cite{Penrose2, Penrose3}.   He also provided a heuristic estimation for the collapse rate, which was similar to the expression suggested by an earlier  collapse model of Diosi \cite{Diosi1, Diosi2}. The DP model is special in that it has a version that is largely parameter-independent. However, this version appears to have been ruled out in recent experiments \cite{PC20}.

The DP model predicts decoherence in the position basis. Other models with the same prediction involve more parameters. They include the   Powers-Persival \cite{PowPer} collapse model from fluctuations of the conformal factor, and the Asprea, Gasbari and Bassi model \cite{AGB} where  particles are under the influence of  a stochastic gravitational field. 

Other versions of fundamental decoherence proceed from the assumption of fluctuations in time / spacetime, leading to an irreducible imprecision in the measuring devices \cite{GPP1, GPP2}, to modified dynamics \cite{Milburn1, Milburn2}, or to modified statistics \cite{Boni}. These models are related to the basic spacetime structure, but they are not explicitly gravitational, and they make strong  assumptions about the behavior of spacetime fluctuations \cite{AnHu08}.  There are also many suggestions about Planck scale processes leading to decoherence. However, it is rather implausible that such conjectural processes would come to dominate over low energy decoherence processes that are adequately described by classical spacetime concepts \cite{AnHu07}.

As mentioned earlier, most models of gravitational decoherence focus on massive particles in the non-relativistic regime. This is partly because of the simplicity of calculations in this regime---gravitational effects are reduced to Newtonian gravity---and partly because the strength of  gravitational decoherence effects is expected to increase with the particle mass. 

There are few quantitative predictions for gravitational decoherence of photons---for a recent toy model with a finite number of degrees of freedom, see, Ref. \cite{XuBle20}. While photons in the optical and ultraviolet spectrum interact much more weakly with gravity than massive particles, it is possible to design quantum optical interference experiments with very long baselines, which could lead to the identification of   minute decoherence effects.    For example, in the proposed Deep Space Quantum Link program \cite{DSQL}, baselines for optical experiments may reach up to $10^5$ km. 

To the best of our knowledge, the only quantitative model for gravitational decoherence of photons is the so-called {\em event formalism} of Ralph, Milburn and Downes \cite{event1, event2}, which predicts disentanglement of photons in certain regimes. However, the underlying  physics is rather implausible from the perspective of gravity theory. The event formalism relies on the presence of closed-timelike curves \cite{deutsch} without accompanying quantum gravity effects. This strongly contradicts the well-motivated  {\em chronology protection conjecture} \cite{HawkingCP}. This conjecture asserts that the laws of physics, including quantum phenomena, do not allow for the appearance of closed time-like curves. It is well supported at the level of semi-classical gravity \cite{KRW}, strongly suggesting that closed timelike curves emerge only from quantum gravity effects (like Wheeler's spacetime foam \cite{Wheeler}), if at all.

A key motivation for this work is the necessity to develop gravitational decoherence models for photons that are based on plausible gravitational physics. To this end, we generalize the ABH model of gravitational decoherence \cite{AnHu13, Blencowe} to photons. The ABH model is the most conservative gravitational decoherence model in that it assumes maximal validity of QFT and General Relativity. The agents of decoherence are linearized stochastic fluctuations of the gravitational  field, and their action upon matter is described in terms of the theory of open quantum systems \cite{Davies, BrePe07}. In Ref. \cite{Blencowe}, the origin of the fluctuations is cosmological. In Ref. \cite{AnHu13}, fluctuations originate from the emergence of classical spacetime geometry as a thermodynamic variable; as such they are of thermodynamic / statistical origin. Another calculation in this vein is in Ref. \cite{OnWa1}, and the results also apply to photons; however, the evolution equation derived there is a pre-master equation that does not lend itself to quantitative calculations. 
For other calculations / models on the decoherence effects from graviton reservoirs, see Refs. \cite{Ana96, Hab00, HaKl1, RLLNJ}.

We must emphasize, that unlike  many models of fundamental decoherence, we are not looking for a fundamental  mechanism that induces the quantum-to-classical transition, and we do not view ABH-type decoherence as a solution to the quantum measurement problem \cite{Sch, Ana02}. We are looking for non-unitary channels based on plausible physics, of fundamental origin, that could be manifested in interferometry experiments.

We start our anslysis from the action describing the interaction of the electromagnetic field with linearized gravity. Then, we perform a 3+1 decomposition, and pass to the Hamiltonian formalism. We quantize the system, and we trace out the degrees of freedom of the gravitational field. Thus, we construct the second-order master equation \cite{Davies, BrePe07}. We identify the term responsible for noise and decoherence for general EM field states. This term is largely insensitive to the choice of initial state of the gravitational perturbations: it depends only on the state's behavior in the deep infrared. For  a Gaussian initial state, this results into a decoherence term that depends only on a single free parameter, an effective {\em noise temperature} $\Theta$. The latter is not necessarily a physical temperature, hence, it is not constrained by any microscopic considerations. In principle, it can take values  larger than the Planck temperature $T_P \simeq 1.4 \times 10^{32}  K$.

Then, we show that values of $\Theta$ near $T_P$ lead to appreciable loss of   interferometric visibility, for states with a large number of photons and for baselines relevant to deep-space experiments. We also discuss other conceivable ways of testing the decoherence effect in photons.  

\medskip

The plan of this paper is the following. In Sec. 2, we analyse the classical EM field coupled to linearized gravity. In Sec. 3, we quantize the system, construct the master equation for the EM field, and we identify the term responsible for decoherence and noise. In Sec. 4, we analyse decoherence phenomena and possible experiments with photons in which they can be distinguished. In Sec. 5, we summarize our results.

Unless stated otherwise, we use units with $ c = \hbar = k_B =1$. The metric's signature is $(-,+,+,+)$.

\section{The classical Hamiltonian for the EM field interacting with weak gravity}

The starting point of our analysis, is the classical action functional describing the EM field $A_{\mu}$ and the spacetime metric $g_{\mu \nu}$. The action is
\begin{eqnarray}
S = \frac{1}{2\kappa} \int{d^4 x \sqrt{-g} \, R} - \frac{1}{4} \int{d^4 x \sqrt{-g} \, F^{\mu\nu} F_{\mu\nu}} \, ,
\end{eqnarray}
where $ g $ is the determinant of the metric $ g_{\mu\nu} $, $ R $ is the Ricci scalar, $ F_{\mu\nu} = \nabla_\mu A_\nu - \nabla_\nu A_\mu $ is the EM tensor and we have defined the constant $ \kappa = 8 \pi G $.

We assume that the spacetime manifold $M$ is of topology $ \mathbbm{R}  \times \Sigma$, and that the spacetime $(M, g)$ is globally hyperbolic. We introduce a spacelike foliation, with time coordinate $t$ and spatial coordinates $ x^i $. After a 3+1 decomposition, the action becomes
\begin{eqnarray}
S =& \frac{1}{2\kappa} \int dt \, d^3 x \, N\sqrt{h} \left(\tensor[^{(3)}]{R}{} + K^{ij} K_{ij} - K^2\right) - \frac{1}{4} \int dt \, d^3 x  \, N\sqrt{h} \, h^{ij}  \nonumber \\
&\times \bigg(-\frac{2}{N^2}(\dot A_i+\partial_i \Phi)(\dot A_j+\partial_j \Phi) - \frac{4N^k}{N^2}(\dot A_i+\partial_i \Phi)(\partial_j A_k - \partial_k A_j) \nonumber \\
&+ \Big(h^{mn} - \frac{2N^m N^n}{N^2}\Big)(\partial_i A_m - \partial_m A_i)(\partial_j A_n - \partial_n A_j)\bigg) \, ,
\end{eqnarray}
where the dot signifies the Lie derivative with respect to the vector field $ \partial / \partial t $, $ A_i $ is the vector EM potential, $ \Phi = - A_0 $ is the scalar EM potential, $ N $ is the lapse function, $ N^i $ is the shift vector;  $ \tensor[^{(3)}]{R}{} $ and $ K_{ij} = (\dot h _{ij} - \nabla_i N_j -\nabla_j N_i)/(2N) $ is the Ricci scalar and the extrinsic curvature on $ \Sigma $, respectively.

The conjugate momenta are
\begin{eqnarray}
\quad p^{ij} &= \frac{\delta\mathcal{L}}{\delta\dot{h}_{ij}} = \frac{\sqrt{h}}{2\kappa}(K^{ij} - h^{ij}K) \, ,
\end{eqnarray}
\begin{eqnarray}\label{electr}
-E^i &= \frac{\delta\mathcal{L}}{\delta\dot A _i} = \frac{\sqrt{h} \, h^{ij}}{N}\Big(\dot A_j+\partial_j \Phi - N^k(\partial_k A_j - \partial_j A_k)\Big) \, ,
\end{eqnarray}
\begin{eqnarray}
\quad \pi_0 &= \frac{\delta\mathcal{L}}{\delta\dot{N}} = 0 \, , \\
\quad \pi_i &= \frac{\delta\mathcal{L}}{\delta\dot{N^i}} = 0 \, , \\
\quad \pi_\Phi &= \frac{\delta\mathcal{L}}{\delta\dot{\Phi}} = 0 \, ,
\end{eqnarray}
defined by the Lagrangian density $ \mathcal{L} $ associated to  $ S $.

We perform a Legendre-transform on the Lagrangian to obtain the Hamiltonian
\begin{eqnarray}
H =& \int d^3 x \, \bigg(\frac{2 \, \kappa \, N}{\sqrt h}\big(h_{im}h_{jn} - \frac{1}{2} \, h_{ij}h_{mn} \big) \, p^{ij}p^{mn} -2 \, N_j \, \partial_i p^{ij} \nonumber \\
 &- \frac{N\sqrt{h}}{2\kappa}\, \tensor[^{(3)}]{R}{} +  \frac{N \, h_{ij}}{2\sqrt h} \, E^i E^j+N \sqrt h \, h^{ij}h^{mn} \, \partial_{[i}A_{m]} \, \partial_{[j}A_{n]} \nonumber \\
&- 2\, N^i E^j \, \partial_{[i}A_{j]} - \Phi \, \partial_i E^i + \pi_0 \dot N + \pi_i \dot N^i + \pi_\Phi \dot \Phi \bigg).
\end{eqnarray}

Next, we take the weak-field limit for gravity,
\begin{eqnarray}
h_{ij} &= \delta_{ij} + 2 \, \kappa \, \gamma_{ij} \, , \\
N &= 1+\kappa \, L \, ,\\
N_i &= \kappa \, L_i \, ,
\end{eqnarray}
 keeping only up to first order terms of $ \kappa $. By defining $ B^i = \varepsilon^{ijk}\partial_j A_k $, we obtain
\begin{eqnarray}
H =& \int d^3 x \left(\frac{E^2+B^2}{2} \right) +2\kappa \int d^3 x \left(p_{ij} \, p^{ij}-\frac{p^2}{2} \right) \nonumber \\
&+\kappa \int d^3 x \left(\frac{1}{2} \, \partial_i \gamma_{jk} \, \partial^i \gamma^{jk}-\partial_i \gamma_{jk} \, \partial^k \gamma^{ji}+\partial_i \gamma \, \partial_j \gamma^{ij}-\frac{1}{2} \, \partial_i \gamma \, \partial^i \gamma \right) \nonumber \\
&+\kappa \int d^3 x \, \gamma_{ij} \left( E^iE^j+B^iB^j-\delta^{ij}\frac{E^2+B^2}{2} \right) \nonumber \\
&-\int d^3 x \, \Phi \, \partial_i E^i  +\kappa \int d^3 x \, L \left( - \partial^i\partial^j \gamma_{ij} + \partial^2 \gamma +\frac{E^2+B^2}{2} \right) \\
&-\kappa \int d^3 x \, L^i \left(  2 \, \partial^j p_{ij} +\varepsilon_{ijk}E^j B^k  \right) + \int d^3 x \left( \kappa \pi_0 \dot L + \kappa \pi_i \dot L^i + \pi_\Phi \dot \Phi \right) \, . \nonumber
\end{eqnarray}
Here, the first two rows are the free Hamiltonians of the EM field and of the gravity perturbations, the third row is the interaction Hamiltonian, the fourth and fifth rows are the constraints.

Next, we undertake the constraint analysis of the system. The primary constraints are $ \pi_0 = \pi_i = \pi_\Phi = 0 $, while the  secondary (first class) constraints are
\begin{eqnarray}
C_0 &= \partial^i\partial^j \gamma_{ij} - \partial^2 \gamma - \frac{E^2+B^2}{2} = 0 \, , \\
C_i &= -2\partial^j p_{ij} - \varepsilon_{ijk}E^j B^k = 0 \, ,
\end{eqnarray}
\begin{eqnarray}\label{constr_phi}
C_\Phi &= \partial_i E^i = 0 \, .
\end{eqnarray}

The combination
\begin{eqnarray}
C(\lambda_0, \lambda_i, \lambda_{\Phi}) = \int d^3 x \left[\lambda_0(x) C_0(x) + \lambda^{i}(x) C_{i}(x) + \lambda_{\Phi}(x) C_{\Phi}(x)\right]
\end{eqnarray}
generates the infinitesimal gauge transformations
\begin{eqnarray}\label{diff1}
\delta \gamma_{ij} = \frac{1}{2\kappa}\left( \partial_i \lambda_j + \partial_j \lambda_i \right) \, ,
\end{eqnarray}
\begin{eqnarray}\label{diff2}
\delta p^{ij} = \frac{1}{2\kappa}\left( \delta^{ij} \, \partial^2 \lambda_0 - \partial^i \partial^j \lambda_0 \right) \, ,
\end{eqnarray}
\begin{eqnarray}\label{gauge1}
\delta A_i  = \lambda_0 \, E_i + \lambda^j \left( \partial_j A_i - \partial_i A_j \right) + \partial_i \lambda_\Phi \, ,
\end{eqnarray}
\begin{eqnarray}\label{gauge2}
\delta E^i = - \partial_j \left( \lambda_0 \, \left( \partial^i A^j - \partial^j A^i \right) \right) + \partial_j \left( E^i \, \lambda^j - E^j \, \lambda^i \right) \, .
\end{eqnarray}

For completeness, we evaluate the associated finite transformations in Appendix A. These results may be of significance in a path-integral approach to this system.
In the Appendix, we prove that 
the fields transform according to space and time reparameterizations together with rotations and scaling of the coordinates. The diffeomorphisms associated to the relations (\ref{diff1}) and (\ref{diff2}) conserve the foliation $ \Sigma \times \mathbbm{R} $ of the spacetime manifold, and are not simply the diffeomorfisms encountered in GR. 

The transformations (\ref{diff1}---\ref{diff2}) imply 
that the longitudinal part $ \gamma^{(L)}_{ij} $ and the transverse trace $ p^{(T)} $ are pure gauge. To quantize,  we have to fix the gauge. Even if gauge fixing is, in general, not necessary for quantizing, in this case we are forced to select a gauge condition that preserves the Lorentzian foliation intoduced in the 3+1 decomposition. The reason is that a QFT in Minkowski spacetime is defined only with respect to a Lorentz frame, so that it carries a representation of the Poincar\'{e} group. In Ref. \cite{AnHu13}, it is explained that this gauge dependence is unavoidable, and that it is related to an ambiguity inherent in  quantum treatments of gravity that has been pointed out by Penrose \cite{Penrose2}.

After imposing the gauge condition $ \gamma^{(L)}_{ij} = p^{(T)} = 0 $ and the Coulomb gauge $ \partial_i A^i = 0 $, we can solve the secondary constraints to get the final form of our classical Hamiltonian
\begin{eqnarray}
H =& \int{d^3 x \, \frac{ E^2_{(T)} + B^2 }{ 2 } } + 2\kappa \int{d^3 x \left(p^{(TT)}_{ij}p^{ij}_{(TT)} + \frac{1}{4}\, \partial_i \gamma^{(TT)}_{jk}\partial^i \gamma^{jk}_{(TT)} \right) } \nonumber \\
&+ \kappa  \int{d^3 x \, \gamma^{(TT)}_{ij} \mathcal{T}^{ij}_{(TT)} } - \frac{\kappa \, V}{4} \, , \label{Hamclas}
\end{eqnarray}
where we defined
\begin{eqnarray}
\mathcal{H} &=& \frac{E^2+B^2}{2} \, , \\
\mathcal{P}_i &=& \varepsilon_{ijk}E^j B^k \, , \\
\mathcal{T}^{ij} &=& E^i E^j + B^i B^j - \delta^{ij} \frac{E^2 + B^2}{2} \, , \\
V &=& \int d^3 x \, d^3 x^\prime \, \frac{   \mathcal{H}(\mathbf{x}) \, \mathcal{H}(\mathbf{x^\prime}) - 4 \, \mathcal{P}^{(T)}_i (\mathbf{x}) \,  \mathcal{P}_{(T)}^i(\mathbf{x^\prime}) - \mathcal{P}^{(L)}_i(\mathbf{x}) \, \mathcal{P}_{(L)}^i(\mathbf{x^\prime})   }{4 \pi \, |\mathbf{x} - \mathbf{x^\prime}|}
\end{eqnarray}
and by $ (TT) $ we denote the transverse-traceless part.

\section{ Master equation for the EM field}

\subsection{Quantization}
We proceed by standard perturbative quantization. We express the fields in terms of (i) creation and annihilation operators for photons, $\hat{a}_{\mathbf p , \lambda}, \hat{a}_{\mathbf p , \lambda}^{\dagger}$, ; and (ii) creation and annihilation operators $\hat{b}_{\mathbf k , \sigma}, \hat{b}_{\mathbf k , \sigma}^\dagger$ for the gravitational perturbations . The indices $\lambda$ and $\sigma$ stand for polarization. 

We drop the indices $T$ and $TT$, and write the fields as
\begin{eqnarray}
\hat{A}^n (\mathbf x ) = \int{\frac{d^3 p}{(2\pi)^3}\frac{1}{\sqrt{2 \omega_{\mathbf p}}}\displaystyle\sum_{\lambda=\pm 1}{\left( \epsilon^n_{(\lambda)}(\mathbf{p}) \hat{a}_{\mathbf{p}, \lambda} e^{i \mathbf{p} \cdot \mathbf{x}} + \epsilon^{n*}_{(\lambda)}(\mathbf{p}) \hat{a}_{\mathbf{p}, \lambda}^\dagger e^{- i \mathbf{p} \cdot \mathbf{x} } \right)}} \, , \\
\hat{\gamma}^{mn} (\mathbf x ) = \frac{1}{\sqrt{\kappa}}\int{\frac{d^3 k}{(2\pi)^3}\frac{1}{\sqrt{2 \omega_{\mathbf k}}}\displaystyle\sum_{\sigma=\pm 1}{\left( \epsilon^{mn}_{(\sigma)}(\mathbf k ) \hat{b}_{\mathbf k , \sigma} e^{i {\mathbf k}\cdot{\mathbf x}} + \epsilon^{mn*}_{(\sigma)}(\mathbf k ) \hat{b}_{\mathbf k , \sigma}^\dagger e^{- i {\mathbf k}\cdot{\mathbf x}} \right)}} \, , \nonumber \\
\\
\hat{E}^n (\mathbf x ) = \, i \int{\frac{d^3 p}{(2\pi)^3}\sqrt{\frac{\omega_\mathbf{p} }{ 2 } } \displaystyle\sum_{\lambda=\pm 1}{\left( \epsilon^n_{(\lambda)}(\mathbf{p} ) \hat{a}_{\mathbf{p} , \lambda} e^{i \mathbf{p} \cdot \mathbf{x} } - \epsilon^{n*}_{(\lambda)}(\mathbf{p} ) \hat{a}_{\mathbf{p} , \lambda}^\dagger e^{- i \mathbf{p} \cdot \mathbf{x} } \right)}} \, , \\
\hat{p}^{mn} (\mathbf x ) = -\frac{i}{2\sqrt{\kappa}}\int{\frac{d^3 k}{(2\pi)^3}\sqrt{\frac{\omega_{\mathbf k}}{2}}\displaystyle\sum_{\sigma=\pm 1}{\left( \epsilon^{mn}_{(\sigma)}(\mathbf k ) \hat{b}_{\mathbf k , \sigma} e^{i {\mathbf k}\cdot{\mathbf x}} - \epsilon^{mn*}_{(\sigma)}(\mathbf k ) \hat{b}_{\mathbf k , \sigma}^\dagger e^{- i {\mathbf k}\cdot{\mathbf x}} \right)}} \, . \nonumber \\
\end{eqnarray}
We choose circular polarization, so that the relation $ \epsilon^{mn}_{(\sigma)}(\mathbf k ) = \epsilon^{m}_{(\sigma)}(\mathbf k ) \epsilon^{n}_{(\sigma)}(\mathbf k ) $ holds between the polarization tensors of the gravitons and the polarization vectors of the photons.

Then, the quantum version of the Hamiltonian (\ref{Hamclas}) is
\begin{eqnarray}
\hat{H} =& \hat{H}_{EM}-\frac{\kappa}{4}\, \hat{V}+\hat{H}_g+\hat{H}_I \, ,
\end{eqnarray}
where $ \hat{H}_{EM} = \int \frac{d^3 p}{(2\pi)^3} \sum_{\lambda} \omega_{\mathbf p} \, \hat{a}_{\mathbf p , \lambda}^\dagger \, \hat{a}_{\mathbf p , \lambda } $ and $ \hat{H}_g = \int \frac{d^3 k}{(2\pi)^3} \sum_{\sigma} \omega_{\mathbf k} \, \hat{b}_{\mathbf k , \sigma}^\dagger \, \hat{b}_{\mathbf k , \sigma} $ are the Hamiltonians for the free EM field and the free graviton field respectively. The interaction Hamiltonian $\hat{H}_I$ is
\begin{eqnarray}
\hat{H}_I =& \sqrt{\frac{\kappa}{2}}\int \frac{d^3 k}{(2\pi)^3} \frac{1}{\sqrt{\omega_\mathbf{k}}} \displaystyle\sum_{\sigma=\pm 1} \left( \hat{b}_{\mathbf{k},\sigma} \, \hat{J}^\dagger_\sigma (\mathbf k ) + \hat{b}_{\mathbf{k},\sigma}^\dagger \, \hat{J}_\sigma (\mathbf k ) \right) \, ,
\end{eqnarray}
where we  defined
\begin{eqnarray}
\hat{J}_\sigma (\mathbf k ) &= \epsilon^{(\sigma)*}_{mn}(\mathbf k ) \int d^3 x \, \hat{\mathcal{T}}_{(TT)}^{mn} (\mathbf x )\,  e^{-i\mathbf{k}\cdot\mathbf{x}} \\
&= \int \frac{d^3 p}{(2\pi)^3} \displaystyle\sum_{\lambda=\pm 1} \Big( 2 \, L^{(+,+)}_{(\sigma,\lambda)}(\mathbf k , \mathbf p )\, \hat{a}_{\mathbf{p},\lambda}^\dagger \, \hat{a}_{\mathbf{p}+\mathbf{k},\lambda} \nonumber \\
&\qquad -  L^{(-,-)}_{(\sigma,\lambda)}(\mathbf k , \mathbf p ) \, \hat{a}_{\mathbf{p},-\lambda} \, \hat{a}_{\mathbf{k}-\mathbf{p},\lambda} - L^{(+,-)}_{(\sigma,\lambda)}(\mathbf k , \mathbf p ) \, \hat{a}_{\mathbf{p},\lambda}^\dagger \, \hat{a}_{-\mathbf{p}-\mathbf{k},-\lambda}^\dagger \Big)
\end{eqnarray}
together with the functions
\begin{eqnarray}
L^{(z,w)}_{(\sigma,\lambda)}(\mathbf k , \mathbf p ) =& \sqrt{\omega_\mathbf{p} \, \omega_{\mathbf{p}+ z \mathbf{k} } } \left( \bm{\epsilon}_{(-\sigma)} (\mathbf{k}) \cdot \bm{\epsilon}_{(-\lambda)}(\mathbf{p}) \right) \left( \bm{\epsilon}_{(-\sigma)} (\mathbf{k}) \cdot \bm{\epsilon}_{(w \lambda)}(\mathbf{p}+ z \, \mathbf{k}) \right) \nonumber \\
\end{eqnarray}
and $ z, w = \pm 1 $. We note that $ \hat{J}_\sigma^\dagger (\mathbf k )  = \hat{J}_\sigma ( - \mathbf k ) $. 

The self-interaction operator $ \hat{V} $ can be written as
\begin{eqnarray}\label{selfint}
\hat{V} =& \int \frac{d^3 k}{(2\pi)^3} \left( \frac{\hat{ \mathfrak{h}}^\dagger (\mathbf{k}) \, \hat{\mathfrak{h}}(\mathbf{k}) }{\omega_{\mathbf{k}}^2} - 4 \left(\delta^{nm} - \frac{3 \, k^n k^m}{4\, \omega_{\mathbf{k}}^2} \right) \frac{ \hat{\mathfrak{p}}_n^\dagger (\mathbf{k}) \, \hat{\mathfrak{p}}_m (\mathbf{k})}{\omega_{\mathbf{k}}^2} \right) \nonumber \\
&+ \frac{\Lambda_p^4}{4\pi^2} \int d^3 k \, \frac{ \delta (\mathbf{k}) \, \hat{\mathfrak{h}} (\mathbf{k}) }{\omega_{\mathbf{k}}^2} \, ,
\end{eqnarray}
where $ \Lambda_p $ is a regulator for the UV cutoff for photons and $ \hat{\mathfrak{h}}(\mathbf{k}) $, $ \hat{\mathfrak{p}}^n (\mathbf{k}) $ are the Fourier transforms of the (normal ordered) operators $ :\hat{\mathcal{H}}(\mathbf{x}): $, $ :\hat{\mathcal{P}}^n (\mathbf{x}): $ respectively. Explicitly,
\begin{eqnarray}
\hat{\mathfrak{h}} (\mathbf{k}) =& \int \frac{d^3 p}{(2\pi)^3} \displaystyle\sum_{\lambda=\pm 1} \bigg( \sqrt{\omega_\mathbf{p} \, \omega_\mathbf{p+k}} \, \bm{\epsilon}_{(-\lambda)}(\mathbf{p}) \cdot \bm{\epsilon}_{(\lambda)}(\mathbf{p+k}) \, \hat{a}_{\mathbf{p},\lambda}^\dagger \hat{a}_{\mathbf{p+k},\lambda} \nonumber \\
&\qquad - \frac{\sqrt{\omega_\mathbf{p} \, \omega_\mathbf{-p+k}}}{2} \, \bm{\epsilon}_{(-\lambda)}(\mathbf{p}) \cdot \bm{\epsilon}_{(\lambda)}(\mathbf{-p+k}) \, \hat{a}_{\mathbf{p},-\lambda} \hat{a}_{\mathbf{-p+k},\lambda} \nonumber \\
&\qquad - \frac{\sqrt{\omega_\mathbf{p} \, \omega_\mathbf{-p-k}}}{2} \, \bm{\epsilon}_{(-\lambda)}(\mathbf{p}) \cdot \bm{\epsilon}_{(\lambda)}(\mathbf{-p-k}) \, \hat{a}_{\mathbf{p},\lambda}^\dagger \hat{a}_{\mathbf{-p-k},-\lambda}^\dagger \bigg) \\
\hat{\mathfrak{p}}_n (\mathbf{k}) =&  -i \, \varepsilon_{njl} \, \int \frac{d^3 p}{(2\pi)^3} \displaystyle\sum_{\lambda=\pm 1} \lambda \bigg( \sqrt{\omega_\mathbf{p} \, \omega_\mathbf{p+k}} \, \epsilon^j_{(-\lambda)}(\mathbf{p}) \, \epsilon^l_{(\lambda)}(\mathbf{p+k}) \, \hat{a}_{\mathbf{p},\lambda}^\dagger \hat{a}_{\mathbf{p+k},\lambda} \, , \nonumber \\
&\qquad - \frac{\sqrt{\omega_\mathbf{p} \, \omega_\mathbf{-p+k}}}{2} \, \epsilon^j_{(-\lambda)}(\mathbf{p}) \, \epsilon^l_{(\lambda)}(\mathbf{-p+k}) \, \hat{a}_{\mathbf{p},-\lambda} \hat{a}_{\mathbf{-p+k},\lambda} \nonumber \\
&\qquad - \frac{\sqrt{\omega_\mathbf{p} \, \omega_\mathbf{-p-k}}}{2} \, \epsilon^j_{(-\lambda)}(\mathbf{p}) \, \epsilon^l_{(\lambda)}(\mathbf{-p-k}) \, \hat{a}_{\mathbf{p},\lambda}^\dagger \hat{a}_{\mathbf{-p-k},-\lambda}^\dagger \bigg) \, .
\end{eqnarray}
The last term in equation (33) corresponds to self-interactions of photons through the graviton vacuum state. Since this term takes the form
\begin{eqnarray}
\frac{\Lambda_p^4}{4\pi^2} \int d^3 k \, \frac{ \delta (\mathbf{k}) \, \hat{\mathfrak{h}} (\mathbf{k}) }{\omega_{\mathbf{k}}^2} &= \frac{\Lambda_p^4}{4\pi^2 \, \epsilon_g^2} \hat{H}_{EM}
\end{eqnarray}
it corresponds to a renormalization of the photon wavefunction, where $ \epsilon_g $ is a virtual mass of the graviton. Hence, we are justified in ignoring this term.

\subsection{The reduced EM field dynamics}

Next, we trace out the graviton degrees of freedom treating them as an environment. We assume the weak coupling limit and that the bath is  stationary. The latter condition means that     $ \hat{\rho}_{sys} (t) \approx \hat{\rho} (t) \otimes \hat{\rho}_g $, where $ \hat{\rho}_{sys} (t) $ is the density matrix of the full system, $ \hat{\rho} (t) $ is the density matrix of the photons and $ \hat{\rho}_g  $ is the density matrix of the gravitational perturbations. With these assumptions, the effective dynamics of the EM field are well described by the second-order master equation.  

The second-order  master equation can be derived in different ways. One way involves the van Hove limit, i.e., the limit $\kappa \rightarrow 0$  with $\kappa^2 t$  fixed \cite{Davies}. A different approach involves the successive use of the Born approximation, the Markov approximation and the Rotating Wave Approximation (RWA) \cite{BrePe07}. In this paper, we follow the latter approach, as we want to investigate the structure of the master equation prior to the implementation of the RWA.
 
 The second-order master equation for the EM-field degrees of freedom takes the form
\begin{eqnarray}\label{2OME}
\frac{d\hat{\rho}}{dt} &= -i \, [ \hat{H}_{EM} - \frac{\kappa}{4} \, \hat{V}, \hat{\rho} ] + \mathcal{D}[ \hat{\rho} ] \, ,
\end{eqnarray}
where $\mathcal{D}[ \hat{\rho} ]$ is a superoperator that contains all information of the graviton bath. In general, $\mathcal{D}[ \hat{\rho} ]$ splits in two terms
\begin{eqnarray}\label{superop_split}
\mathcal{D}[ \hat{\rho} ] = \mathcal{D}_{LS}[ \hat{\rho} ] + \mathcal{D}_{Lin}[ \hat{\rho} ] \, .
\end{eqnarray}
The term $\mathcal{D}_{LS}[ \hat{\rho} ]$ is of the form $-i \frac{\kappa}{2} [\hat{V}_{LS}, \hat{\rho}]$, i.e., it leads to the addition of a term $\hat{V}_{LS}$ to the total Hamiltonian. Here, LS stands for Lamb-shift, because  this term implements a field-induced correction to the energy levels of the system. 
The term $\mathcal{D}_{Lin}[ \hat{\rho} ]$ is of the Lindblad form
\begin{eqnarray}
\mathcal{D}_{Lin}[ \hat{\rho} ]= \sum_a c_a \left[\hat{L}_a\hat{\rho}\hat{L}^{\dagger}_a - \frac{1}{2}\hat{L}^{\dagger}_a\hat{L}_a \hat{\rho} - \frac{1}{2} \hat{\rho}\hat{L}^{\dagger}_a\hat{L}_a \right],
\end{eqnarray}
where $\hat{L}_a$ are operators on the Fock space of the EM field.
 
 The explicit form of $\mathcal{D}[ \hat{\rho} ]$ depends on the initial state of the graviton bath $\hat{\rho}_g$. Different physical assumptions about the nature of the bath correspond to different choices for $\hat{\rho}_g$. 
 \begin{enumerate}
     \item In perturbative quantum gravity,  Minkowski spacetime corresponds to the ground state of a quantum gravity theory, hence, to the graviton ground state in the linearized approximation. If we choose $\hat{\rho}_g$ as the graviton vacuum, then we can show that $\mathcal{D}_{Lin}[ \hat{\rho} ]$ vanishes. Vacuum graviton fluctuations induce no decoherence, except possibly for transient terms at early times that are not captured by the second-order master equation---see, for example, Ref. \cite{Ana96}. The term $\mathcal{D}_{LS}[ \hat{\rho} ]$ renormalizes the EM Hamiltonian. The renormalization terms are different from the standard analysis of perturbative quantum gravity, as the initial state is factorized and hence, not the true vacuum.
     
     \item We can assume a thermal bath of gravitons, possibly of cosmological origin, as in Ref. \cite{Blencowe}. The natural temperature here would be that of the cosmic microwave background. In general, this temperature is too small to cause appreciable decoherence.
     
     \item One may also choose a quantum state that mimics the effects of a classical stochastic background, for, example a Gaussian stochastic process \cite{ReJa}. This background may be caused, for example, by all rotating neutron stars in the galaxy. As the source of the noise is of astrophysical origin, and of known physics, its magnitude can be approximately estimated.
     
     \item In Ref. \cite{AnHu13}, Anastopoulos and Hu proposed a different physical origin for the fluctuations of the gravitational field. If the notion of a classical spacetime is emergent from a quantum gravity theory at the thermodynamic level of description, then Minkowkski spacetime describes a macrostate of a quantum gravity theory---rather than a quantum state as in perturbative quantum gravity. Hence, the emergence of classical spacetime should be accompanied by thermodynamic fluctuations which are classicalized and much stronger than any quantum ones \cite{Hu09}. The size of the fluctuations cannot be fixed by known physics, it emerges from a more fundamental quantum theory of gravity. In \cite{AnHu13}, these fluctuations were referred to as {\em spacetime textures}.
     
 \end{enumerate}
 
 As in Ref. \cite{AnHu13}, we will choose for $\hat{\rho}_g$ a thermal state at an effective temperature $\Theta$. This choice is not restrictive, because  it turns out that the Lindblad term  $\mathcal{D}_{Lin}[ \hat{\rho} ]$   depends only on the  behavior of the state at the limit $\omega \rightarrow 0$. Hence, any  state with the same structure at the deep-infrared---including states that mimic classical noise---leads to the same non-unitary dynamics, hence, to the same terms for dissipation and noise. In contrast, the Lamb-shift terms strongly depend on the choice of $\hat{\rho}_g$, and so does the renormalization of the Hamiltonian. 

Then, the non-unitary term depends only on a single parameter, the effective temperature $\Theta$. This parameter interpolates between the vacuum of perturbative quantum gravity ($\Theta = 0$), the low value $\Theta \simeq 2.7 K$ for conjectured cosmological graviton perturbations, and to very large values that correspond to classical noise. Note that in the latter case, $\Theta$ is best interpreted as noise temperature\footnote{The power carried by a noise is given by $P \sim\Theta \Delta \omega$, where $\Delta \omega$ is the band-width of the noise and $\Theta$ the noise temperature.}, and not as physical temperature. Hence, $\Theta$ needs not be bounded by above by the natural upper bound to physical temperatures,namely, Planck temperature. Possible cosmological constraints to $\Theta$ are currently under consideration.

With the above choice as the initial state, we evaluate the superoperator $\mathcal{D}[ \hat{\rho} ]$, subject to the Born and Markov approximations --but prior to the RWA \cite{BrePe07}-- as
\begin{eqnarray}\label{superop}
\mathcal{D}[ \hat{\rho} ] &= \frac{\kappa}{2} \int \frac{d^3 k}{(2\pi)^3} \displaystyle\sum_{\sigma=\pm 1} \frac{ \coth \left( \frac{\omega_{\mathbf{k}} }{2 \Theta} \right)}{\omega_{\mathbf{k}} } \left( \Big[ \hat{J}_\sigma (\mathbf{k}) , [ \hat{\rho} , \hat{\widetilde{J}^\dagger_{\sigma}} (\mathbf{k} ) ] \Big] + \Big[ \hat{J}^\dagger_\sigma (\mathbf{k}) , [ \hat{\rho} , \hat{\widetilde{J}_{\sigma}} (\mathbf{k} ) ] \Big] \right) \nonumber \\
&+ \frac{\kappa}{2} \int \frac{d^3 k}{(2\pi)^3} \displaystyle\sum_{\sigma=\pm 1} \frac{1}{\omega_{\mathbf{k}} } \left( \Big[ \hat{J}_\sigma (\mathbf{k}) , \{ \hat{\rho} , \hat{\widetilde{J}^\dagger_{\sigma}} (\mathbf{k} ) \} \Big] - \Big[ \hat{J}^\dagger_\sigma (\mathbf{k}) , \{ \hat{\rho} , \hat{\widetilde{J}_{\sigma}} (\mathbf{k} ) \} \Big] \right)
\end{eqnarray}
in terms of the operators
\begin{eqnarray}\label{Jtilde}
\hat{\widetilde{J}_{\sigma}} (\mathbf{k} ) &=& \int \frac{d^3 p}{(2\pi)^3} \displaystyle\sum_{\lambda=\pm 1}  \bigg( 2 \, L^{(+,+)}_{(\sigma,\lambda)}(\mathbf k , \mathbf p ) \, \hat{a}_{\mathbf{p},\lambda}^\dagger \, \hat{a}_{\mathbf{p+k},\lambda} \, f ( \omega_\mathbf{k} + \omega_\mathbf{p} - \omega_\mathbf{p+k} ) \nonumber \\
&-& L^{(-,-)}_{(\sigma,\lambda)}(\mathbf k , \mathbf p ) \, \hat{a}_{\mathbf{p},-\lambda} \, \hat{a}_{\mathbf{k-p},\lambda} \, f(\omega_\mathbf{k} - \omega_\mathbf{p} - \omega_\mathbf{p-k} ) \nonumber \\
&-& L^{(+,-)}_{(\sigma,\lambda)}(\mathbf k , \mathbf p ) \, \hat{a}_{\mathbf{p},\lambda}^\dagger \, \hat{a}_{\mathbf{-p-k},-\lambda}^\dagger  \, f ( \omega_\mathbf{k} + \omega_\mathbf{p} + \omega_\mathbf{p+k} ) \bigg) \, ,
\end{eqnarray}
where
\begin{eqnarray}\label{HeaviFour}
f (\omega) = \int_{-\infty}^{\infty} ds \, u(s) \, e^{-i \omega s} = \pi \, \delta(\omega) - i \, \textrm{p.v.} \left( \frac{1}{\omega} \right)
\end{eqnarray}
and $ u(s) $ is the Heaviside step function and ``p.v."  stands  for Cauchy principal value. 

A similar  equation has been derived in Ref. \cite{OnWa1}.The main difference is that the evolution equation of \cite{OnWa1} is derived by a cumulant expansion rather than by the Born-Markov equation. This leads to a difference in details, for example, in the definition of their analogue to the operator $\hat{J}_{\sigma}$, and there is also a difference by an overall multiplicatie factor   $-\frac{1}{2}$ on the superoperator. Note, however, that the derivation of a completely positive master equation requires additional approximations, like the Rotation Wave Approximation, that follows. Such approximations have not been efected in Ref. \cite{OnWa1}, with the result that their evolution equation does not provide any simple quantitative predictions.

\subsection{The Rotating Wave Approximation}

We define the operators $\hat{J}^a_{(\sigma,\lambda)} (\mathbf{k} , \mathbf{p} )$ for $a = 1, 2, 3$, as
\begin{eqnarray}
\hat{J}^1_{(\sigma,\lambda)} (\mathbf{k} , \mathbf{p} ) &= 2 \, L^{(+,+)}_{(\sigma,\lambda)}(\mathbf k , \mathbf p ) \, \hat{a}_{\mathbf{p},\lambda}^\dagger \, \hat{a}_{\mathbf{p+k},\lambda} \, , \\
\hat{J}^2_{(\sigma,\lambda)} (\mathbf{k} , \mathbf{p} ) &= - L^{(-,-)}_{(\sigma,\lambda)}(\mathbf k , \mathbf p ) \, \hat{a}_{\mathbf{p},-\lambda} \, \hat{a}_{\mathbf{k-p},\lambda} \, , \\
\hat{J}^3_{(\sigma,\lambda)} (\mathbf{k} , \mathbf{p} ) &= - L^{(+,-)}_{(\sigma,\lambda)}(\mathbf k , \mathbf p ) \, \hat{a}_{\mathbf{p},\lambda}^\dagger \, \hat{a}_{\mathbf{-p-k},-\lambda}^\dagger \, .
\end{eqnarray}
Then, we define the functions $\omega_a (\mathbf{k} , \mathbf{p} ) $ for $a = 1, 2, 3$, as
\begin{eqnarray}
\omega_1 (\mathbf{k} , \mathbf{p} ) &= \omega_\mathbf{p} - \omega_\mathbf{p+k} \, , \\
\omega_2 (\mathbf{k} , \mathbf{p} ) &= - \omega_\mathbf{p} - \omega_\mathbf{p-k} \, , \\
\omega_3 (\mathbf{k} , \mathbf{p} ) &= \omega_\mathbf{p} + \omega_\mathbf{p+k} \, .
\end{eqnarray}

With these definitions,
\begin{eqnarray}
\hat{J}_{\sigma} (\mathbf{k}) &= \int \frac{d^3 p}{(2\pi)^3} \displaystyle\sum_{\lambda ,a}  \hat{J}^a_{(\sigma,\lambda)} (\mathbf{k} , \mathbf{p} ) \, , \\
\hat{\widetilde{J}_{\sigma}} (\mathbf{k} ) &= \int \frac{d^3 p}{(2\pi)^3} \displaystyle\sum_{\lambda ,a}  \hat{J}^a_{(\sigma,\lambda)} (\mathbf{k} , \mathbf{p} ) f( \omega_\mathbf{k} + \omega_a (\mathbf{k} , \mathbf{p} ) ) \, .
\end{eqnarray}

The RWA involves the suppression of rapidly oscillating terms from $\mathcal{D}[ \hat{\rho} ]$  in the interaction representation---see, Ref. \cite{BrePe07}. 
In this system, the oscillating phases are of the form $e^{i[\omega_a (\mathbf{k} , \mathbf{p} ) - \omega_b (\mathbf{k} , \mathbf{p^\prime})]t}$ for different $a, b$ and $\mathbf{p}, \mathbf{p}'$.

The RWA is the condition that we keep only terms in which the oscillating factor vanishes---see, for example Sec. 3.3 in Ref. \cite{BrePe07}. This means that we only keep terms 
\begin{eqnarray}\label{cond}
\omega_a (\mathbf{k} , \mathbf{p} ) = \omega_b (\mathbf{k} , \mathbf{p^\prime} )
\end{eqnarray}
in the superoperator (\ref{superop}).

After application of the condition (\ref{cond}), the superoperator of (\ref{superop}) becomes
\begin{eqnarray}
\mathcal{D}[ \hat{\rho} ] = \frac{\kappa \, \Theta}{4\pi} \left( \left[ \hat{P}_{ij} , \left[ \hat{\rho} , \hat{P}^{ij} \right] \right] - \frac{1}{3} \left[ \hat{H} , \left[ \hat{\rho} , \hat{H} \right] \right] \right) - i \, \frac{\kappa}{2} \left[ \hat{V}_{LS} , \hat{\rho} \right] \, ,
\end{eqnarray}
where we have defined the operators
\begin{eqnarray}
\hat{P}^{ij} = \int \frac{d^3 p}{(2\pi)^3} \displaystyle\sum_{\lambda = \pm 1} \frac{p^i \, p^j }{\omega_\mathbf{p}} \hat{a}_{\mathbf{p},\lambda}^\dagger \, \hat{a}_{\mathbf{p},\lambda}
\end{eqnarray}
and
\begin{eqnarray}\label{VLS}
\hat{V}_{LS} =& \sum_{a=1}^3 \int \frac{d^3 k \, d^3 p \, d^3 p^\prime}{(2\pi)^9} \sum_{\sigma,\lambda,\lambda^\prime} \textrm{p.v.}\left( \frac{ 1 }{ \omega_\mathbf{k} + \omega_a (\mathbf{k} , \mathbf{p} ) } \right) \nonumber \\
&\qquad \quad \times \Bigg( \frac{ \coth \left( \frac{ \omega_\mathbf{k} }{2 \Theta} \right) }{ \omega_\mathbf{k} } \left[ \hat{J}^a_{(\sigma,\lambda)} (\mathbf{k} , \mathbf{p} ) , \hat{J}^{a \, \dagger}_{(\sigma,\lambda^\prime)} (\mathbf{k} , \mathbf{p^\prime} ) \right] \nonumber \\
&\qquad \qquad \quad - \left. \frac{ 1 }{ \omega_\mathbf{k} } \left\{ \hat{J}^a_{(\sigma,\lambda)} (\mathbf{k} , \mathbf{p} ) , \hat{J}^{a \, \dagger}_{(\sigma,\lambda^\prime)} (\mathbf{k} , \mathbf{p^\prime} ) \right\} \Bigg) \right|_{\omega_a (\mathbf{k} , \mathbf{p} ) = \omega_a (\mathbf{k} , \mathbf{p^\prime} )}
\end{eqnarray}
and we have dropped the index $ EM $ of the Hamiltonian for the free EM field.

The Lindblad part of $\mathcal{D}[ \hat{\rho} ]$ can be brought in the diagonal form
\begin{eqnarray}\label{Lind1a}
\mathcal{D}_{Lin} [ \hat{\rho} ] = \frac{\kappa \, \Theta}{2 \pi} \left( \hat{C}_{ij}  \hat{\rho} \, \hat{C}^{ij}  - \frac{1}{2} \{ \hat{\rho} , \hat{C}_{ij}  \hat{C}^{ij}  \} \right) \, ,
\end{eqnarray}
where 
\begin{eqnarray}
\hat{C}_{ij} = \hat{P}_{ij} - \frac{1}{3} \, \delta_{ij} \, \hat{H} \, . 
\end{eqnarray}

\subsection{The weak-RWA}
The RWA guarantees that the master equation is completely positive. However, it is not a necessary condition; meaningful evolution equations can be obtained with weaker assumptions \cite{FlHu}. Furthermore, the RWA might misrepresent some features like entanglement generation and destruction \cite{FCAH}. 

With an eye to possible future applications to entangled states, we also investigated a weaker version of the RWA, to which we refer as the weak-RWA. The detailed calculation is in the Appendix B. We found  a master equation with a Lindblad term,
\begin{eqnarray}\label{wRWA}
\mathcal{D}_{Lin}[ \hat{\rho} ] &= \frac{\kappa \, \Theta}{2 \pi} \left( \hat{C}_{ij}^{(1)} \hat{\rho} \, \hat{C}^{ij}_{(1)} - \frac{1}{2} \{ \hat{\rho} , \hat{C}_{ij}^{(1)} \hat{C}^{ij}_{(1)} \} \right) \nonumber \\
&-\frac{\kappa \, \Theta}{2 \pi} \left( \hat{C}_{ij}^{(2)} \hat{\rho} \, \hat{C}^{ij}_{(2)} - \frac{1}{2} \{ \hat{\rho} , \hat{C}_{ij}^{(2)} \hat{C}^{ij}_{(2)} \} \right) \, ,
\end{eqnarray}
where
\begin{eqnarray}
\hat{C}_{ij}^{(1)} = \hat{P}_{ij} - \frac{1}{3} \, \delta_{ij} \, \hat{H} + \frac{ \hat{\Pi}_{ij}^\dagger + \hat{\Pi}_{ij} }{2} \, , \\
\hat{C}_{ij}^{(2)} = \frac{ \hat{\Pi}_{ij}^\dagger + \hat{\Pi}_{ij} }{2} \, .
\end{eqnarray}

The difference from Eq. (\ref{Lind1a}) is found in the presence of the operators
\begin{eqnarray}
\hat{\Pi}^{ij} = \int \frac{d^3 p}{(2\pi)^3} \displaystyle\sum_{\lambda = \pm 1} \omega_\mathbf{p} \, \epsilon^i_{(\lambda)}(\mathbf{p}) \, \epsilon^j_{(\lambda)}(\mathbf{p}) \, \hat{a}_{\mathbf{p},\lambda} \, \hat{a}_{-\mathbf{p},-\lambda} \, ,
\end{eqnarray}
which involve a pair of annihilation operators with opposite momentum. They contribute non-negligibly only for states that have support on a subspace of momenta that includes opposite vectors. They can be ignored for all other states, including ones that describe linearly propagating waves.

\subsection{Single-photon states}

Next, we restrict the master equation (\ref{2OME}) to the single-photon subspace. Hence, we obtain the evolution equation at the level of individual photons. 

To this end, we assume a density matrix of the form
\begin{eqnarray}
\hat{\rho}_t =& \displaystyle\sum_{\lambda,\lambda^\prime} \int \frac{d^3 p}{(2\pi)^3} \, \frac{d^3 p^\prime}{(2\pi)^3} \, \rho (\mathbf{p},\lambda;\mathbf{p^\prime},\lambda^\prime ; t) \, \hat{a}^\dagger_{\mathbf{p},\lambda} | 0 \rangle \langle 0 | \hat{a}_{\mathbf{p^\prime},\lambda^\prime} \, ,
\end{eqnarray}
where $ \rho (\mathbf{p},\lambda;\mathbf{p^\prime},\lambda^\prime ; t) $ is a single-particle density matrix, and we only keep terms in the master equation that conserve the single particle subspace. We work in the momentum representation. Reinserting the constant $ \kappa = 8 \pi G $, we write the master equation for a single photon
\begin{eqnarray}\label{result}
\dot{ \hat{ \rho }_t } &= -i \, [ \hat{H }_R , \hat{\rho}_t ] + 2 \, G \, \Theta \left( \delta^{in} \delta^{jm} -\frac{1}{3} \delta^{ij} \delta^{nm} \right) \left[ \frac{\hat{p}_i \, \hat{p}_j}{\hat{p}_0} , \left[ \hat{\rho}_t , \frac{\hat{p}_n \, \hat{p}_m}{\hat{p}_0} \right] \right] \, ,
\end{eqnarray}
where $ \hat{H}_R = \hat{H} - \frac{\kappa}{4} \hat{V} + \frac{\kappa}{2} \hat{V}_{LS}  $ is the renormalized Hamiltonian of the photon (see Appendix C), and $ \hat{p}_0 = \sqrt{ \hat{p}_i \, \hat{p}^i } $.

\section{Decoherence / dephasing phenomena}

First, we note that the open-system dynamics do not distinguish between photons of different polarizations. This is because the initial state is invariant under exchange of graviton polarization. While this is a natural assumption, it is by no means guaranteed: it is  conceivable that the spacetime textures of Ref. \cite{AnHu13} are chiral. 

In the present model, the open-system dynamics do not affect polarization, so we drop its dependence from the density matrix. 

For motion in one direction,  
\begin{eqnarray}\label{master_1D}
\mathcal{D}_{Lin}[ \hat{\rho} ] = \frac{4 \, G \, \Theta}{3} \left[ \hat{H} , \left[ \hat{\rho} , \hat{H} \right] \right] \, .
\end{eqnarray}
Hence, for a large class of states in which the gravitational self-interaction of light is negligible, we obtain an one-dimensional master equation of the form (reinserting $\hbar$, $c$ and $k_B$)
\begin{eqnarray}\label{1dim}
\dot{ \hat{ \rho }_t } &= -\frac{i}{\hbar} \, \left[ \hat{H} , \hat{\rho}_t \right] - \frac{\tau}{2 \, \hbar^2} \left[ \hat{H} , \left[ \hat{H} , \hat{\rho}_t \right] \right] \, ,
\end{eqnarray}
where 
\begin{eqnarray}
\tau = \frac{8 \, G \, k_B \, \Theta}{3 \, c^5}
\end{eqnarray}
is a constant of dimension time.

 A master equation like (\ref{1dim}) appears in models by Milburn \cite{Milburn1}, Adler \cite{Adler}, Diosi \cite{Diosi} and Breuer et al \cite{Breuer} where $\tau$ is obtained from postulated stochastic fluctuations of time, discreteness of time, stochastic fluctuations of the metric,  or even stochastic fluctuations of $\hbar$. Note, however, that the papers above either do not refer specifically to photons or they refer to non-relativistic particles.  
 
 In the models above, the value of $\tau$ is not fixed, but the natural candidate is the Planck-time $\tau_P \sim 10^{-43}s$.  In Ref. \cite{AnHu13}, it is argued that  open system dynamics that cause decoherence are not constrained by the dynamics of classical GR, hence, the choice 
 $\tau = \tau_P$ is not in any sense natural. Here, $\tau$ is determined by  $\Theta$ which is a free parameter. Note that if $\tau \sim \tau_P$, then also $\Theta \sim T_P$.

When the Lindblad operator coincides with the Hamiltonian operator,  time evolution leads to diagonalization of the density matrix in the energy basis \cite{Ana96}. So,  Eq. (\ref{1dim})   can be  solved in   this eigenstate basis
\begin{eqnarray}\label{1dim2}
\rho_t (E,E^\prime) &= \exp \left[ -\frac{i}{\hbar} \, (E- E^\prime) \, t - \frac{\tau}{2 \, \hbar^2} (E - E^\prime)^2 \, t \right] \rho_0 (E,E^\prime) \, .
\end{eqnarray}
Eq. (\ref{1dim2}), then, implies a  decoherence rate:
\begin{eqnarray}
\Gamma = \frac{4 \, G \, k_B \, \Theta}{3 \, \hbar^2 \, c^5} (\Delta E)^2 \, .
\end{eqnarray}
Since the diagonalization of the density matrix is only approximate, there are probably other observables with respect to which the density matrix becomes approximately diagonal. This is not the case, however, for the `position' matrix elements, obtained by Fourier transforming the momentum matrix elements of the density matrix.

For setups not restricted in one dimension, we must return to Eq. (\ref{result}), which in the momentum representation can be solved as
\begin{eqnarray}\label{3dim2}
\rho_t (\mathbf{p},\mathbf{p^\prime}) &= \exp \left[ -\frac{i \, c}{\hbar} \, (p- p^\prime) \, t - \frac{c^2 \tau}{2 \, \hbar^2} \left( (p - p^\prime)^2 + 3 \, p \, p^\prime \sin^2 \gamma \right) \, t \right] \rho_0 \, ,
\end{eqnarray}
where $ \gamma $ is the angle between $ \mathbf{p} $ and $ \mathbf{p^\prime} $. Consider an experiment involving photons of energy $E$ that propagate distance $L$ along different arms of an interferometer,   which are placed at right angles, so that $\gamma = \pi/2$.

Suppose there is a small difference $\delta L << L$ of optical length along the two arms of the interferometer. By Eq. (\ref{3dim2}), the amplitude of the interference fringes, proportional to $ \cos^2 [\frac{p \, \delta L}{2 \hbar}] $, will be suppressed by the exponential $\nu(t) = \exp[-\frac{3 \, \tau \, E^2 t}{2 \, \hbar^2 }] $. 

Exponential suppression factors in interferometry are commonly written as $\exp[- \frac{1}{2} (\Delta \Phi)^2]$, because such terms can be  obtained from the stochastic average $\langle e^{i\delta \phi}\rangle$ of a stochastic phase change $\delta \phi$ due to the presence of a decohering environment---$(\Delta \Phi)^2 = \langle \delta \phi^2\rangle$ with respect to a Gaussian stochastic process. In this notation, 
\begin{eqnarray}
(\Delta \Phi)^2 = \frac{3 \, \tau \, E^2 t}{ \, \hbar^2 } = \frac{8 \, G \, k_B \, \Theta E^2 L}{\hbar^2 \, c^6} \, .
\end{eqnarray}

Let us assume that $  E$ is of the order of $1eV$. For $\Theta \sim 3 K$, we need $L \sim 10^{53} m$ (much larger than the visible Universe) for total loss of  visibility, $(\Delta \Phi)^2 \sim 1$. For $\Theta \sim T_P$, total loss of  visibility requires an interferometer arm of the size of the galaxy ($L \sim 10^{21}m$). Obviously, total loss of  visibility in interference experiments with light is undetectable.

However, gravitational wave interferometry works with  phase changes of the order of $\delta \phi \sim 10^{-9}$ \cite{Thrn}. Such accuracy is possible because the error in the determination of the phase is of the order of $\bar{N}^{-1/2}$, where $\bar{N}$ is the average number of photons in a quantum state. Hence, a small loss of visibility may be detectable for EM-field states with a large number of photons, like coherent states of large amplitude. The master equation (\ref{2OME}) derived here also applies to such states. 

Proposed deep space experiments may involve interferometric arm lengths $L \sim 10^5km$. For $\Theta \sim T_P$, and $E $ of the order of $1 eV$, this implies a loss of coherence $\Delta \Phi \sim 10^{-8}$. In principle, this is discernible with EM-field states with $\bar{N} > 10^{16}$.

The quadratic dependence on energy implies that dephasing is significantly stronger at high frequencies. For interferometry at the extreme UV, $ (\Delta \Phi)^2 $ may increase by four orders of magnitude or more.

Furthermore, it may be possible to identify the effects of gravitational decoherence on photons arriving from remote sources, as is common in quantum gravity phenomenology---, see, Ref. \cite{QGP1, QGP2} and references therein. Suppose there is a  process that creates photons described by a density matrix with non-zero off-diagonal matrix elements $\rho(E, E')$. Then, by Eq. (\ref{1dim2}), the off-diagonal elements are suppressed exponentially with time $t$, hence, with the distance $L$ from the source. This distance-dependent suppression of coherence can be identified in probability distribution of photonic observables that do not commute with the Hamiltonian. The prime candidate for such observables is the (relativistic) time-of-arrival \cite{Werner, Leon, AnSav12, AnSav19}.
 
This reasoning will not work with ordinary light from stellar sources, which is well modeled by a thermal initial state that is diagonal in the energy basis. However, terms beyond the RWA, as in Eq. (\ref{wRWA}) could lead to the appearance of non-diagonal elements out of a diagonal initial state, and thereby lead to novel effects beyond the ones considered here.

Of course, the tiny effect of gravitational decoherence on photons may be hidden due to the presence of stronger, non-gravitational sources of decoherence. That may also be an issue for experiments in deep space, where the interaction of light particles with charges (through the interaction with solar wind, for example) can be seen to have big effects \cite{Kief}. Shielding photons from other sources of decoherence would be an important challenge in any experimental setup.

    \section{Conclusions}
We constructed the master equation for the quantum EM field interacting with a stochastic graviton bath. Thus, we generalized the ABH model for gravitational decoherence to photons. We also showed that deep space experiments can place strong bounds on the value of the free parameter $\Theta$ that describes the strength of gravitational  perturbations.  

The methodology that we employed here can be straightforwardly generalized to other systems, for example, neutrinos. It can also be used in order to develop more complex models of gravitational decoherence, for example, ones involving chiral gravitational perturbations.

Finally, we point out that the derivation of a second-order master equation, as effected here, is the most reliable method for identifying the effect of different graviton reservoirs on a quantum system, and thereby, to look for characteristic quantum signatures in the noise, as proposed in Ref. \cite{PWZ}.

\section*{Acknowledgements}
We would like to thank Bei-Lok Hu for many discussions on the topic.

 \begin{appendix}

\section{ Finite gauge transformations}
In this section, we evaluate explicitly the finite gauge transformations for the classical EM field interacting with gravity perturbations. This calculation is included for completeness, but also for its relevance to any path-integral treatment of the systems studied here.

\subsection{A model calculation for a single variable}
First, we show how to obtain finite gauge transformations for the case of functions of a single variable.
Consider a function $ \phi (x) $ with infinitesimal transformation of the form
\begin{eqnarray}
\delta \phi (x) = \alpha (x) \frac{d \phi (x)}{dx} + \beta (x) \phi (x) + \gamma (x) \, .
\end{eqnarray}
Its finite transformation will be
\begin{eqnarray}\label{full_transform}
\phi^\prime (x) = e^{ \alpha (x) \frac{d}{dx} + \beta (x) } \phi (x) + \left( e^{ \alpha (x) \frac{d}{dx} + \beta (x) } - 1 \right) \left( \alpha (x) \frac{d}{dx} + \beta (x) \right)^{-1} \gamma (x) \, , \nonumber \\
\quad
\end{eqnarray}
where the operator $ \left( \alpha (x) \frac{d}{dx} + \beta (x) \right)^{-1} $ is the inverse of $ \alpha (x) \frac{d}{dx} + \beta (x) $, or in other words, acting on a function $ \gamma (x) $ gives the solution of the differential equation
\begin{eqnarray}
& \left( \alpha (x) \frac{d}{dx} + \beta (x) \right) y(x) = \gamma (x) \\
\Rightarrow \qquad & y(x) = e^{-\int_0^x \frac{\beta (v)}{\alpha (v)} dv} \left( \int_0^x \frac{\gamma (u)}{\alpha (u)} e^{\int_0^u \frac{\beta (v)}{\alpha (v)} dv} du +c \right)
\end{eqnarray}
(assuming of course $ \alpha (x) \neq 0 $). Thus, we have
\begin{eqnarray}
\phi^\prime (x) =& e^{ \alpha (x) \frac{d}{dx} + \beta (x) } \phi (x) \nonumber \\
&+ \left( e^{ \alpha (x) \frac{d}{dx} + \beta (x) } - 1 \right) e^{-\int_0^x \frac{\beta (v)}{\alpha (v)} dv} \left( \int_0^x \frac{\gamma (u)}{\alpha (u)} e^{\int_0^u \frac{\beta (v)}{\alpha (v)} dv} du + c \right) \, .
\end{eqnarray}
It can be shown that
\begin{eqnarray}
e^{\alpha (x) \frac{d}{dx} + \beta (x) } = \exp{\left( \int_0^1 e^{t \alpha (x) \frac{d}{dx}} \beta (x) e^{-t \alpha (x) \frac{d}{dx}} dt \right) } e^{\alpha (x) \frac{d}{dx}} \, ,
\end{eqnarray}
where the first exponent is simply the function
\begin{eqnarray}
\int_0^1 e^{t \alpha (x) \frac{d}{dx}} \beta (x) e^{-t \alpha (x) \frac{d}{dx}} dt &= \left(e^{\alpha (x) \frac{d}{dx}} - 1 \right) \left( \alpha (x) \frac{d}{dx} \right)^{-1} \beta (x) \nonumber \\
&= \left(e^{\alpha (x) \frac{d}{dx}} - 1 \right) \int_0^x \frac{\beta (v)}{\alpha(v)} dv \, .
\end{eqnarray}
The operator $ e^{\alpha (x) \frac{d}{dx}} $ acting on a function $ \psi (x) $ gives
\begin{eqnarray}
e^{\alpha (x) \frac{d}{dx}} \psi (x) = \psi ( T(x) ) \, ,
\end{eqnarray}
with $ T(x) $ a function that satisfies the relation
\begin{eqnarray}
\int_x^{T(x)} \frac{dv}{\alpha (v)} = 1 \, .
\end{eqnarray}
This lets us write the finite transformation as
\begin{eqnarray}
\phi^\prime (x) = e^{\int_x^{T(x)} \frac{\beta (v)}{\alpha (v)} dv} \phi (T(x)) + \int_x^{T(x)} \frac{\gamma (u)}{\alpha (u)} e^{\int_x^u \frac{\beta (v)}{\alpha (v)} dv} du \, .
\end{eqnarray}

In the special case where $\alpha (x) = 0$, we can compute the above transformation from eq. (\ref{full_transform}):
\begin{eqnarray}
\phi^\prime (x) =& e^{\beta (x)} \phi (x) + \left(e^{\beta (x)} - 1 \right) \frac{\gamma (x)}{\beta (x)} \, .
\end{eqnarray}

\subsection{Multi-variable case}

Generalization to higher dimensions is straight forward. Assuming an infinitesimal transformation, written here in vector format
\begin{eqnarray}
\delta \phi (\mathbf{x}) = \bm{\alpha} (\mathbf{x}) \cdot \nabla \phi (\mathbf{x}) + \beta (\mathbf{x}) \phi (\mathbf{x}) + \gamma (\mathbf{x}) \, ,
\end{eqnarray}
the finite transformation of $ \phi (\mathbf{x}) $ can then be given by
\begin{eqnarray}\label{full_transform_final}
\phi^\prime (\mathbf{x}) = e^{B (\mathbf{x},\mathbf{x^\prime})} \phi (\mathbf{x^\prime}) + \Gamma (\mathbf{x},\mathbf{x^\prime})
\end{eqnarray}
with
\begin{eqnarray}
B (\mathbf{x},\mathbf{x^\prime}) = \int_\mathbf{x}^\mathbf{x^\prime} \beta (\mathbf{r}) \, \frac{\bm{\alpha} (\mathbf{r}) \cdot d\mathbf{r} }{|\bm{\alpha} (\mathbf{r})|^2} \, , \\
\Gamma (\mathbf{x},\mathbf{x^\prime}) = \int_\mathbf{x}^\mathbf{x^\prime} e^{B (\mathbf{x},\mathbf{r})} \, \gamma (\mathbf{r}) \, \frac{\bm{\alpha} (\mathbf{r}) \cdot d\mathbf{r} }{|\bm{\alpha} (\mathbf{r})|^2}
\end{eqnarray}
and $ \mathbf{x^\prime} $ is a transformation of the coordinates that satisfies
\begin{eqnarray}
\int_\mathbf{x}^\mathbf{x^\prime} \frac{\bm{\alpha} (\mathbf{r}) \cdot d\mathbf{r}}{|\bm{\alpha} (\mathbf{r})|^2} = 1 \, .
\end{eqnarray}
In all the above curvilinear integrals we must integrate along a curve, such that, on every point of the curve, the vector $ \bm{\alpha} (\mathbf{x}) $ is tangent (i.e. along a flow line of $ \bm{\alpha} (\mathbf{x}) $), at least in the general case where $ \nabla \times \left( \bm{\alpha} (\mathbf{x}) / |\bm{\alpha} (\mathbf{x})|^2 \right) \neq 0 $.

Of interest is also the case where the above functions do not depend on $ \mathbf{x} $ (meaning $ \bm{\alpha} (\mathbf{x}) = \bm{\alpha} $, $ \beta (\mathbf{x}) = \beta $ and $ \gamma (\mathbf{x}) = \gamma $):
\begin{eqnarray}
\phi^\prime (\mathbf{x}) =& e^{\beta} \phi (\mathbf{x} + \bm{\alpha}) + \frac{e^{\beta} - 1}{\beta} \gamma \, .
\end{eqnarray}

\subsection{Finite transformations of the EM tensor}
Next we apply the results of the above analysis to the infinitesimal gauge transformations of Sec. 2. We employ the relations
\begin{eqnarray}
\partial_0 A_i = \{A_i,H\} =& - E_i + \partial_i A_0 + \mathcal{O}(\kappa) \, , \\
\partial_0 E^i = \{E^i,H\} =& \partial_j \left( \partial^i A^j - \partial^j A^i \right) + \mathcal{O}(\kappa) \, ,
\end{eqnarray}
while discarding terms of order $ \kappa $ and above, and applying the constraint (\ref{constr_phi}). Then,  the transformations (\ref{gauge1}), (\ref{gauge2}) can be rewritten as
\begin{eqnarray}
A^\prime_i = A_i + \lambda^\mu \, \partial_\mu A_i + A_\mu \partial_i \lambda^\mu + \partial_i f \, , \\
E^{\prime \, i} = E^i + \lambda^\mu \, \partial_\mu E^i + E^i \partial_j \lambda^j - E^j \partial_j \lambda^i + F^{ij} \partial_j \lambda^0 \, ,
\end{eqnarray}
where $ \lambda^0 = - \lambda_0 $, $ f = \lambda_\Phi - \lambda^\mu A_\mu $ and $ F^{ij} = \partial^i A^j - \partial^j A^i $.

Eq. (\ref{diff1}) shows us that these tranformations take us out of the linearized theory, so for the primed fields the raising and lowering of indices is performed by multiplying with the new metric
\begin{eqnarray}
h^\prime_{ij} = \delta_{ij} + \partial_i \lambda_j + \partial_j \lambda_i + \mathcal{O}(\kappa) \, .
\end{eqnarray}
This leads us to compute the transformed components $ F^\prime_{i0} = \partial_i A^\prime_0 - \partial_0 A^\prime_i $ of the EM tensor (for infinitesimal $ \lambda^\mu $) through eq. (\ref{electr}):
\begin{eqnarray}
F^\prime_{i0} = \frac{ N h^\prime_{ij} E^{\prime j} }{ \sqrt{h} } + \mathcal{O}(\kappa) = \left( \delta_{ij} + \partial_i \lambda_j + \partial_j \lambda_i - 2 \delta_{ij} \partial_k \lambda^k \right) E^{\prime j} + \mathcal{O}(\kappa) \, , \nonumber \\
\quad
\end{eqnarray}
so ignoring again the terms of order $ \kappa $
\begin{eqnarray}\label{comp1}
F^\prime_{i0} = F_{i0} + \lambda^\mu \, \partial_\mu F_{i0} + F_{j0} \partial_i \lambda^j - F_{i0} \partial_j \lambda^j + F_{ij} \partial^j \lambda^0 \, ,
\end{eqnarray}
\begin{eqnarray}\label{comp2}
F^\prime_{ij} = F_{ij} + \lambda^\mu \, \partial_\mu F_{ij} + F_{i \mu} \partial_j \lambda^\mu + F_{\mu j} \partial_i \lambda^\mu
\end{eqnarray}
and by combining eq. (\ref{comp1}) and (\ref{comp2}):
\begin{eqnarray}\label{inf_diff_EM}
F^\prime_{\mu \nu} = F_{\mu \nu} + \lambda^\sigma \, \partial_\sigma F_{\mu \nu} + T\indices{_\mu^\sigma} F_{\sigma \nu} + T\indices{_\nu^\sigma} F_{\mu \sigma} \, ,
\end{eqnarray}
where
\begin{eqnarray}
T\indices{_0^0} &= -\partial_i \lambda^i \, , \\
T\indices{_i^0} &= \partial_i \lambda^0 \, , \\
T\indices{_0^i} &= \partial^i \lambda^0 \, , \\
T\indices{_i^j} &= \partial_i \lambda^j \, .
\end{eqnarray}

The associated finite transformation can be acquired by exponentiation of eq. (\ref{inf_diff_EM}):
\begin{eqnarray}\label{trans_EM}
F^\prime_{\mu \nu}(x) &= \exp \left( \delta_\mu^\sigma \, \delta_\nu^\rho \, \lambda^\alpha \, \partial_\alpha + \delta_\mu^\sigma \, T\indices{_\nu^\rho} + \delta_\nu^\rho \, T\indices{_\mu^\sigma} \right) F_{\sigma \rho}(x) \nonumber \\
&= \Lambda\indices{_\mu^\sigma} \, \Lambda\indices{_\nu^\rho} \, F_{\sigma \rho}(x^\prime) \, ,
\end{eqnarray}
with
\begin{eqnarray}
\Lambda\indices{_\mu^\nu} = \exp \left( \int_x^{x^\prime} T\indices{_\mu^\nu} (\xi) \frac{ \lambda^\alpha(\xi) \, d\xi_\alpha }{ \lambda^\beta(\xi) \, \lambda_\beta(\xi) } \right) \, ,
\end{eqnarray}
while the transformed coordinates $ x^{\prime \mu} $ are given by the relation
\begin{eqnarray}
\int_x^{x^\prime} \frac{ \lambda^\alpha(\xi) \, d\xi_\alpha }{ \lambda^\beta(\xi) \, \lambda_\beta(\xi) } = 1
\end{eqnarray}
and integration is performed in the direction of $ \lambda^\mu $.

Imposing the restriction $ \partial_i \lambda_j + \partial_j \lambda_i = 0 $, the above transformations allow us to stay in linearized gravity. In this case, the tensors $ \Lambda\indices{_\mu^\nu} $ are just Lorentz transformations with rapidity vector $ \zeta^i $ and axis-angle vector $ \theta^i $, which are defined as
\begin{eqnarray}
\zeta^i = \int_x^{x^\prime} \partial^i \lambda^0 (\xi) \frac{ \lambda^\alpha(\xi) \, d\xi_\alpha }{ \lambda^\beta(\xi) \, \lambda_\beta(\xi) } \, , \\
\theta^i = - \frac{1}{2} \varepsilon^{ijk} \int_x^{x^\prime} \partial_j \lambda_k (\xi) \frac{ \lambda^\alpha(\xi) \, d\xi_\alpha }{ \lambda^\beta(\xi) \, \lambda_\beta(\xi) } \, .
\end{eqnarray}

In the simplest case where $ \lambda^\mu $  are constant, the finite gauge transformations (\ref{trans_EM}) give us a transformed EM field by the spatial and temporal reparameterizations $ x^{\prime \mu} = x^\mu + \lambda^\mu $
\begin{eqnarray}
F^\prime_{\mu \nu}(x) = F_{\mu \nu}(x+\lambda) \, .
\end{eqnarray}

\section{  The weak-RWA}

In this section, we undertake the investigation of a weaker version of the RWA in the derivation of the master equation. We refer to this version as the {\em weak-RWA}.

The weak-RWA consists in employing the RWA condition (\ref{cond}) only for the functions of Eq. (\ref{HeaviFour}), so that
\begin{eqnarray}\label{weakRWA}
f(\omega_\mathbf{k} + \omega_a (\mathbf{k},\mathbf{p}) ) = f(\omega_\mathbf{k} + \omega_b (\mathbf{k},\mathbf{p^\prime}) ) \, .
\end{eqnarray}
%
We apply  Eq. (\ref{weakRWA}) to the superoperator (\ref{superop}), to obtain
\begin{eqnarray}
\mathcal{D}[ \hat{\rho} ] &= \pi \, \kappa \sum_{a,b} \delta (\omega_\mathbf{k} + \omega_a ) \frac{ \coth \left( \frac{ \omega_\mathbf{k} }{2 \Theta} \right) }{ \omega_\mathbf{k} } \Big( \hat{J}^a \, \hat{\rho} \, \hat{J}^{b \, \dagger} + \hat{J}^{b \, \dagger} \, \hat{\rho} \, \hat{J}^a - \frac{1}{2} \left\{ \hat{\rho} , \left\{ \hat{J}^a , \hat{J}^{b \, \dagger} \right\} \right\} \Big) \nonumber \\
&+ \pi \, \kappa \sum_{a,b} \delta (\omega_\mathbf{k} + \omega_a ) \frac{ 1 }{ \omega_\mathbf{k} } \Big( \hat{J}^a \, \hat{\rho} \, \hat{J}^{b \, \dagger} - \hat{J}^{b \, \dagger} \, \hat{\rho} \, \hat{J}^a + \frac{1}{2} \left\{ \hat{\rho} , \left[ \hat{J}^a , \hat{J}^{b \, \dagger} \right] \right\} \Big) \nonumber \\
&- i \, \frac{\kappa}{2} \left[ \hat{V}_{LS} , \hat{\rho} \right] \, ,
\end{eqnarray}
where we have used the shorthand $ \int \frac{d^3 k \, d^3 p \, d^3 p^\prime}{(2\pi)^9} \sum_{\sigma,\lambda,\lambda^\prime,a,b} \rightarrow \sum_{a,b} $, $ \omega_a (\mathbf{k},\mathbf{p}) \rightarrow \omega_a $, $ \hat{J}^a_{(\sigma,\lambda)} (\mathbf{k} , \mathbf{p} ) \rightarrow \hat{J}^a $, $ \hat{J}^{b \, \dagger}_{(\sigma,\lambda^\prime)} (\mathbf{k} , \mathbf{p^\prime} ) \rightarrow \hat{J}^{b \, \dagger} $, and we defined
\begin{eqnarray}
\hat{V}_{LS} = \sum_{a,b} \textrm{p.v.}\left( \frac{ 1 }{ \omega_\mathbf{k} + \omega_a } \right) \left( \frac{ \coth \left( \frac{ \omega_\mathbf{k} }{2 \Theta} \right) }{ \omega_\mathbf{k} } \left[ \hat{J}^a , \hat{J}^{b \, \dagger} \right] - \frac{ 1 }{ \omega_\mathbf{k} } \left\{ \hat{J}^a , \hat{J}^{b \, \dagger} \right\} \right) \, .
\end{eqnarray}
We can simplify the delta functions $ \delta (\omega_\mathbf{k} + \omega_a ) $ by the following identities
\begin{eqnarray}
\delta ( \omega_\mathbf{k} + \omega_\mathbf{p} - \omega_\mathbf{p+k} ) =& \, \frac{\delta (\omega_\mathbf{k})+\delta (\omega_\mathbf{p})}{1- \cos \gamma_\mathbf{kp}} + \frac{\omega_\mathbf{p} + \omega_\mathbf{k}}{\omega_\mathbf{p} \, \omega_\mathbf{k}} \, \delta (\cos \gamma_\mathbf{kp}-1) \, , \\
\delta ( \omega_\mathbf{k} - \omega_\mathbf{p} - \omega_\mathbf{p-k} ) =& \, \frac{ \delta (\omega_\mathbf{p})}{1- \cos \gamma_\mathbf{kp}} \nonumber \\
&+ \frac{\omega_\mathbf{p} - \omega_\mathbf{k}}{\omega_\mathbf{p} \, \omega_\mathbf{k}} \, u(\omega_\mathbf{p} - \omega_\mathbf{k}) \, \delta (\cos \gamma_\mathbf{kp}-1) \, , \\
\delta ( \omega_\mathbf{k} + \omega_\mathbf{p} + \omega_\mathbf{p+k} ) =& \, 0
\end{eqnarray}
and by the relations
\begin{eqnarray}
L^{(+,+)}_{(\sigma,\lambda)}(\mathbf{k} , \mathbf{p} ) \, \delta (\cos \gamma_\mathbf{kp}-1) &= \, 0 \, , \\
L^{(-,-)}_{(\sigma,\lambda)}(\mathbf{k} , \mathbf{p} ) \, u(\omega_\mathbf{p} - \omega_\mathbf{k}) \, \delta (\cos \gamma_\mathbf{kp}-1) &= \, 0 \, , \\
L^{(z,w)}_{(\sigma,\lambda)}(\mathbf{k} , \mathbf{p} ) \, \delta (\omega_\mathbf{p}) &= \, 0 \, ,
\end{eqnarray}
with $ \gamma_\mathbf{kp} $ the angle between vectors $\mathbf{k}$ and $\mathbf{p}$. 

Then,  the only non vanishing terms in the Linblad part of $ \mathcal{D}[\hat{\rho}] $ are those with $ \delta (\omega_\mathbf{k}) $, so the relevant terms are those with $a=1$. The full RWA condition (\ref{cond}) then for $b=1$ is satisfied identically, while for $b \neq 1$ it is $\omega_\mathbf{p^\prime}=0$. Continuing with the weak-RWA, we have
\begin{eqnarray}
\mathcal{D}_{Lin}& [ \hat{\rho} ] = \frac{\kappa \, \Theta}{2 \pi^2} \int \frac{d^3 p \, d^3 p^\prime}{(2\pi)^6} \displaystyle\sum_{\lambda, \lambda^\prime} \omega_\mathbf{p} \omega_\mathbf{p^\prime} \Bigg( I^{+}_{\lambda,\lambda^\prime}(\mathbf{e}_\mathbf{p},\mathbf{e}_\mathbf{p^\prime}) \Big( \hat{a}_{\mathbf{p},\lambda}^\dagger \, \hat{a}_{\mathbf{p},\lambda} \, \hat{\rho} \, \hat{a}_{\mathbf{p^\prime},\lambda^\prime}^\dagger \, \hat{a}_{\mathbf{p^\prime},\lambda^\prime} \nonumber \\
&+ \hat{a}_{\mathbf{p^\prime},\lambda^\prime}^\dagger \, \hat{a}_{\mathbf{p^\prime},\lambda^\prime} \, \hat{\rho} \, \hat{a}_{\mathbf{p},\lambda}^\dagger \, \hat{a}_{\mathbf{p},\lambda} - \frac{1}{2} \left\{ \hat{\rho} , \left\{ \hat{a}_{\mathbf{p},\lambda}^\dagger \, \hat{a}_{\mathbf{p},\lambda} , \hat{a}_{\mathbf{p^\prime},\lambda^\prime}^\dagger \, \hat{a}_{\mathbf{p^\prime},\lambda^\prime} \right\} \right\} \Big) \nonumber \\
&\qquad \qquad \qquad \qquad -\frac{I^{-}_{\lambda,\lambda^\prime}(\mathbf{e}_\mathbf{p},\mathbf{e}_\mathbf{p^\prime})}{2} \Big( \hat{a}_{\mathbf{p},\lambda}^\dagger \, \hat{a}_{\mathbf{p},\lambda} \, \hat{\rho} \, \hat{a}_{\mathbf{p^\prime},\lambda^\prime} \, \hat{a}_{\mathbf{-p^\prime},-\lambda^\prime} \nonumber \\
&+ \hat{a}_{\mathbf{p^\prime},\lambda^\prime} \, \hat{a}_{\mathbf{-p^\prime},-\lambda^\prime} \, \hat{\rho} \, \hat{a}_{\mathbf{p},\lambda}^\dagger \, \hat{a}_{\mathbf{p},\lambda} - \frac{1}{2} \left\{ \hat{\rho} , \left\{ \hat{a}_{\mathbf{p},\lambda}^\dagger \, \hat{a}_{\mathbf{p},\lambda} , \hat{a}_{\mathbf{p^\prime},\lambda^\prime} \, \hat{a}_{\mathbf{-p^\prime},-\lambda^\prime} \right\} \right\} \Big) \nonumber \\
&\qquad \qquad \qquad \qquad -\frac{I^{-}_{\lambda,-\lambda^\prime}(\mathbf{e}_\mathbf{p},\mathbf{e}_\mathbf{p^\prime})}{2} \Big( \hat{a}_{\mathbf{p},\lambda}^\dagger \, \hat{a}_{\mathbf{p},\lambda} \, \hat{\rho} \, \hat{a}_{\mathbf{p^\prime},\lambda^\prime}^\dagger \, \hat{a}_{\mathbf{-p^\prime},-\lambda^\prime}^\dagger \nonumber \\
&+ \hat{a}_{\mathbf{p^\prime},\lambda^\prime}^\dagger \, \hat{a}_{\mathbf{-p^\prime},-\lambda^\prime}^\dagger \, \hat{\rho} \, \hat{a}_{\mathbf{p},\lambda}^\dagger \, \hat{a}_{\mathbf{p},\lambda} - \frac{1}{2} \left\{ \hat{\rho} , \left\{ \hat{a}_{\mathbf{p},\lambda}^\dagger \, \hat{a}_{\mathbf{p},\lambda} , \hat{a}_{\mathbf{p^\prime},\lambda^\prime}^\dagger \, \hat{a}_{\mathbf{-p^\prime},-\lambda^\prime}^\dagger \right\} \right\} \Big) \Bigg) \, , \nonumber \\
\quad
\end{eqnarray}
where we have defined
\begin{eqnarray}\label{Iint}
I^{\pm}_{\lambda,\lambda^\prime}(\mathbf{e}_\mathbf{p},\mathbf{e}_\mathbf{p^\prime}) = \displaystyle\sum_{\sigma=\pm1} & \int \frac{d \Omega_\mathbf{k}}{1- \cos \gamma_\mathbf{kp}} \left( \bm{\epsilon}_{(\sigma)} (\mathbf{k}) \cdot \bm{\epsilon}_{(\lambda)}(\mathbf{p}) \right) \left( \bm{\epsilon}_{(\sigma)} (\mathbf{k}) \cdot \bm{\epsilon}_{(-\lambda)}(\mathbf{p}) \right) \nonumber \\
&\times \left( \bm{\epsilon}_{(-\sigma)} (\mathbf{k}) \cdot \bm{\epsilon}_{(\lambda^\prime)}(\mathbf{p^\prime}) \right) \left( \bm{\epsilon}_{(-\sigma)} (\mathbf{k}) \cdot \bm{\epsilon}_{(\mp \lambda^\prime)}(\mathbf{p^\prime}) \right) \, ,
\end{eqnarray}
with $ d \Omega_\mathbf{k} = \sin{\theta_\mathbf{k}} \, d \theta_\mathbf{k} \, d \phi_\mathbf{k} $ the solid angle differential of vector $ \mathbf{k} $. Notice that application of the full RWA condition (\ref{cond}) vanishes the terms with $ I^{-}_{\lambda,\pm \lambda^\prime}(\mathbf{e}_\mathbf{p},\mathbf{e}_\mathbf{p^\prime}) $.

Calculating the above integrals we have the results
\begin{eqnarray}
I^{+}_{\lambda,\lambda^\prime}(\mathbf{e}_\mathbf{p},\mathbf{e}_\mathbf{p^\prime}) = \frac{\pi}{2} \left( \frac{\mathbf{p} \cdot \mathbf{p}^\prime }{ \omega_\mathbf{p} \, \omega_\mathbf{p^\prime} } \right)^2 - \frac{\pi}{6} \, , \\
I^{-}_{\lambda,\lambda^\prime}(\mathbf{e}_\mathbf{p},\mathbf{e}_\mathbf{p^\prime}) = - \pi \left( \frac{ \mathbf{p} \cdot \bm{\epsilon}_{( \lambda^\prime )} ( \mathbf{p}^\prime ) }{ \omega_\mathbf{p} } \right)^2 \, .
\end{eqnarray}
Finally, defining the operators
\begin{eqnarray}
\hat{P}^{ij} = \int \frac{d^3 p}{(2\pi)^3} \displaystyle\sum_{\lambda = \pm 1} \frac{p^i \, p^j }{\omega_\mathbf{p}} \hat{a}_{\mathbf{p},\lambda}^\dagger \, \hat{a}_{\mathbf{p},\lambda} \, , \\
\hat{\Pi}^{ij} = \int \frac{d^3 p}{(2\pi)^3} \displaystyle\sum_{\lambda = \pm 1} \omega_\mathbf{p} \, \epsilon^i_{(\lambda)}(\mathbf{p}) \, \epsilon^j_{(\lambda)}(\mathbf{p}) \, \hat{a}_{\mathbf{p},\lambda} \, \hat{a}_{-\mathbf{p},-\lambda} \, ,
\end{eqnarray}
we can express the superoperator (\ref{superop}) as
\begin{eqnarray}\label{result_App}
\mathcal{D}[ \hat{\rho} ] = \frac{\kappa \, \Theta}{4\pi} \left[ \hat{P}_{ij} - \frac{1}{3} \, \delta_{ij} \, \hat{H} + \hat{\Pi}_{ij}^\dagger + \hat{\Pi}_{ij} , \left[ \hat{\rho} , \hat{P}^{ij} \right] \right] - i \, \frac{\kappa}{2} \left[ \hat{V}_{LS} , \hat{\rho} \right] \, .
\end{eqnarray}
%

\section{  The single-photon renormalized Hamiltonian}

Here we calculate the term $ -i [\hat{H}_R , \hat{\rho}] $ with the renormalized Hamiltonian $  \hat{H}_R = \hat{H} - \frac{\kappa}{4} \hat{V} + \frac{\kappa}{2} \hat{V}_{LS} $ for single photon states. 

In the momentum representation, it is sufficient to calculate the (regularized) action of the operator $\hat{H}_R$ on the single photon state $\hat{a}_{\mathbf{p},\lambda}^\dagger |0\rangle$ and keeping only terms that conserve the number of particles.
\begin{eqnarray}
\hat{H}_R \, \hat{a}^\dagger_{\mathbf{p},\lambda} | 0 \rangle = \left( \omega_\mathbf{p} - \frac{\kappa}{4} \omega_{SI} (\omega_\mathbf{p},\Lambda_g) + \frac{\kappa}{2} \omega_{LS} (\omega_\mathbf{p},\Lambda_g,\Theta) \right) \hat{a}^\dagger_{\mathbf{p},\lambda} | 0 \rangle \, .
\end{eqnarray}
We have defined the integrals
\begin{eqnarray}\label{omega_SI}
\omega_{SI} (\omega_\mathbf{p},\Lambda_g) &= 6 \int \frac{d^3 k}{(2\pi)^3} \displaystyle\sum_{z=\pm 1} \frac{\omega_\mathbf{p} \, \omega_{\mathbf{p+k}} }{\omega_\mathbf{k}^4} \Bigg( \frac{\omega_\mathbf{k}^2}{3} \left| \bm{\epsilon}_{(z \lambda)} (\mathbf{p}) \cdot \bm{\epsilon}_{(\lambda)} (\mathbf{p+k}) \right|^2 \\
&\quad + \mathrm{Re} \left[ ( \mathbf{k} \cdot \bm{\epsilon}_{(\lambda)} (\mathbf{p}) ) ( \mathbf{k} \cdot \bm{\epsilon}_{(-z \lambda)} (\mathbf{p+k}) ) ( \bm{\epsilon}_{(-\lambda)} (\mathbf{p}) \cdot \bm{\epsilon}_{(z \lambda)} (\mathbf{p+k}) ) \right] \nonumber \\
&\quad - \frac{1}{2} \left| \mathbf{k} \cdot \bm{\epsilon}_{(\lambda)} (\mathbf{p}) \right|^2 - \frac{1}{2} \left| \mathbf{k} \cdot \bm{\epsilon}_{(\lambda)} (\mathbf{p+k}) \right|^2 - \frac{\omega_\mathbf{k}^2}{6} \Bigg) \, , \nonumber
\end{eqnarray}
\begin{eqnarray}\label{omega_LS}
\omega_{LS} (\omega_\mathbf{p},\Lambda_g,\Theta) &= 4\int \frac{d^3 k}{(2 \pi)^3} \displaystyle\sum_{\sigma = \pm 1} \displaystyle\sum_{z = \pm 1} \displaystyle\sum_{w = \pm 1} \frac{z \, w \coth \left( \frac{ \omega_\mathbf{k} }{2 \Theta} \right) - 1 }{ \omega_\mathbf{k} } \\
&\qquad \times \left| L^{(z,w)}_{(\sigma,z \, \lambda)}(\mathbf k , \mathbf{p} ) \right|^2 \, \textrm{p.v.} \left( \frac{1}{ \omega_\mathbf{k} + z \, \omega_\mathbf{p} - z \, w \, \omega_{\mathbf{p}+z \, \mathbf{k}} } \right) \, , \nonumber
\end{eqnarray}
which correspond to the energy due to self-interaction and to the Lamb-shift effect, respectively. Here we have introduced $\Lambda_g$ as the UV-cutoff regulator for the energy of the graviton, and the integers $ z $ and $ w $ for the purpose of simplifying our formulas. Note that Eq. (\ref{omega_LS}) is the same for either the full RWA, or the weak-RWA of the previous appendix. The integrals of Eqs. (\ref{omega_SI}) and (\ref{omega_LS}) can be computed exactly, but an assumption must be made for the magnitude of the UV-cutoff $\Lambda_g$ relative to the free-field energy of the photon $\omega_\mathbf{p}$.

For $ \Lambda_g < \omega_\mathbf{p} $:
\begin{eqnarray}
\omega_{SI} (\omega_\mathbf{p},\Lambda_g) &= - \frac{ \Lambda_g \, \omega_\mathbf{p}^2 }{\pi^2} - \frac{ 14 \, \Lambda_g^3 }{45 \pi^2} + \frac{ 19 \, \Lambda_g^5 }{ 525 \pi^2 \, \omega_\mathbf{p}^2 } \, ,
\end{eqnarray}
\begin{eqnarray}
\omega_{LS} (\omega_\mathbf{p},\Lambda_g,\Theta) &= - \frac{ \Lambda_g \, \omega_\mathbf{p}^2 }{3 \pi^2} - s (\Lambda_g,\Theta) \, \omega_\mathbf{p} - \frac{ \Lambda_g^3 }{15 \pi^2} - \frac{ 2 \, \Lambda_g^5 }{ 105 \pi^2 \, \omega_\mathbf{p}^2 } \, .
\end{eqnarray}

For $ \Lambda_g > \omega_\mathbf{p} $:
\begin{eqnarray}
\omega_{SI} (\omega_\mathbf{p},\Lambda_g) &= - \frac{ 2 \, \Lambda_g^2 \, \omega_\mathbf{p} }{3 \pi^2} - \frac{ 8 \, \ln \left( \frac{ \Lambda_g }{ \omega_\mathbf{p} } \right) \, \omega_\mathbf{p}^3 }{15 \pi^2} - \frac{ 124 \, \omega_\mathbf{p}^3 }{225 \pi^2} - \frac{ 2 \, \omega_\mathbf{p}^5 }{35 \pi^2 \, \Lambda_g^2} \, ,
\end{eqnarray}
\begin{eqnarray}
\omega_{LS} (\omega_\mathbf{p},\Lambda_g,\Theta) &= - s (\Lambda_g,\Theta) \, \omega_\mathbf{p} - \frac{ \Lambda_g^2 \, \omega_\mathbf{p} }{3 \pi^2} - \frac{ \omega_\mathbf{p}^3 }{15 \pi^2} - \frac{ 2 \, \omega_\mathbf{p}^5 }{105 \pi^2 \, \Lambda_g^2} \, .
\end{eqnarray}

Here we have defined the function
\begin{eqnarray}
s (\Lambda_g,\Theta) &= \frac{ \Lambda_g^2 }{ 3 \pi^2 } + \frac{ 2 \, \Theta^2 }{ 9 } + \frac{ 4 \, \Theta^2 }{ 3 \pi^2 } \left( \frac{\Lambda_g}{\Theta} \ln \left( 1 - e^{-\Lambda_g / \Theta} \right) - \textrm{Li}_2 \left( e^{-\Lambda_g / \Theta} \right) \right) \, . \nonumber \\
\quad
\end{eqnarray}
Notice that $ \Theta $ is proportional to the mean energy of the graviton ($ \langle \omega_\mathbf{k} \rangle = 3 \, \Theta $), so we have $ \Lambda_g >> \Theta $. Taking the limit $ \Lambda_g \rightarrow \infty $ for the terms that converge and dropping the constant terms, we have the regularized Hamiltonian $\hat{H}_R$, in terms of the free Hamiltonian $\hat{H}$.

For $ \Lambda_g < \omega_\mathbf{p} $:
\begin{eqnarray}
\hat{H}_R = \left( 1 - \kappa \left( \frac{ \Lambda_g^2 }{6 \pi^2} + \frac{ \Theta^2 }{ 9 } \right) \right) \hat{H} + \kappa \, \frac{\Lambda_g}{12 \pi^2} \, \hat{H}^2 - \kappa \, \frac{ 13 \, \Lambda_g^5 }{ 700 \pi^2 } \, \hat{H}^{-2} \, .
\end{eqnarray}

For $ \Lambda_g > \omega_\mathbf{p} $:
\begin{eqnarray}
\hat{H}_R = \left( 1 - \kappa \left( \frac{ \Lambda_g^2 }{6 \pi^2} + \frac{ \Theta^2 }{ 9 } \right) \right) \hat{H} + \kappa \left( \frac{2 \, }{15 \pi^2} \ln \left( \frac{ \Lambda_g }{ \hat{H} } \right) + \frac{47}{450 \pi^2} \right) \hat{H}^3 \, .
\end{eqnarray}

The above relations imply that a scaling of the fields $\hat{E}^i \rightarrow \frac{\hat{\widetilde{E}}^i}{\sqrt{Z}}$, $\hat{B}^i \rightarrow \frac{\hat{\widetilde{B}}^i}{\sqrt{Z}}$ and a scaling of the coupling constant $\kappa \rightarrow Z^2 \, \widetilde{\kappa}$ is needed for renormalization, where $ Z = 1 - \kappa \left( \frac{ \Lambda_g^2 }{6 \pi^2} + \frac{ \Theta^2 }{ 9 } \right) $. Substituting these expressions and taking again the limit $ \Lambda_g \rightarrow \infty $ we have the renormalized energy
\begin{eqnarray}\label{Renorm}
\hat{H}_R &= \hat{H} \qquad \qquad \qquad \qquad \qquad \qquad \qquad &\textrm{(for $ \omega_\mathbf{p} < \Lambda_g $),} \\
\hat{H}_R &= \hat{H} + \frac{\kappa \, \xi}{\hat{H}^2} \qquad \qquad \qquad \qquad \qquad \qquad &\textrm{(for $ \omega_\mathbf{p} > \Lambda_g $),}
\end{eqnarray}
where $ \xi $ is a parameter that must be introduced by addition of the relevant counter-term, and could be experimentally measured.


\end{appendix}


\section*{References}
 
 
\begin{thebibliography}{}
\bibitem{AnHu13} C. Anastopoulos and B. L. Hu,  {\em A Master Equation for Gravitational Decoherence: Probing the Textures of Spacetime},  Class. Quant. Grav. 30, 165007 (2013).

\bibitem{MSPB}
W. Marshall, C. Simon, R. Penrose and D. Bouwmeester, {\em Towards Quantum Superpositions of a Mirror}, Phys. Rev. Lett. 91, 130401 (2003).

\bibitem{maqro}R.  Kaltenbaek et al, {\em Macroscopic Quantum Resonators (MAQRO): 2015 Update},  EPJ Quantum Technology 3, 5 (2016); {\em Towards Space-Based Tests of Macroscopic Quantum Physics},    COSP 42: H0-14-18 (2018).

\bibitem{HSMC}B. Helou, B. J. J. Slagmolen, D. E. McClelland, and Y. Chen, {\em LISA pathfinder appreciably constrains collapse models}, Phys. Rev. D95, 084054 (2017).

\bibitem{PC20}S. Donadi, K. Piscicchia, C. Curceanu, et al, {\em Underground Test of Gravity-Related Wave Function Dollapse} Nat. Phys. (2020).https://doi.org/10.1038/s41567-020-1008-4


\bibitem{Blencowe} M. Blencowe, {\em Effective Field Theory Approach to Gravitationally Induced Decoherence}, Phys. Rev. Lett. 111, 021302 (2013).


\bibitem{Sch} M. Schlosshauer, {\em Decoherence and the Quantum-To-Classical Transition}, (Springer, Berlin 2007).

\bibitem{Ana02} C. Anastopoulos, {\em Frequently Asked Questions About Decoherence},    Int. J. Theor. Phys. 41, 1573–1590 (2002).



\bibitem{Karol}  F. Karolyhazy, {\em Gravitation and quantum mechanics of macroscopic objects}, Nuovo Cim. 52, 390 (1966).

\bibitem{Karol2} F. Karolyhazy,  A. Frenkel, and B. Lukács, {\em On the possible role of gravity in
the reduction of the wave function}, in "Quantum Concepts in Space and Time", R. Penrose and C. J. Isham editors, (Oxford,1986, Clarendon Press).

\bibitem{Penrose1} R. Penrose,{\em Gravity and state vector reduction},  in "Quantum Concepts in Space and Time", R. Penrose and C. J. Isham editors, (Oxford,1986, Clarendon Press).

\bibitem{Diosi1} L. Diosi,  {\em A universal master equation for the gravitational violation of quantum mechanics}, Phys. Lett. 120, 377(1987).

\bibitem{Diosi2} L. Diosi,  {\em Models for universal reduction of macroscopic quantum fluctuations},  Phys. Rev. A40, 1165 (1989).

\bibitem{Penrose2} R. Penrose, {\em On gravity's role in quantum state reduction }, Gen. Rel. Grav. 28, 581 (1996). 

\bibitem{Penrose3} R. Penrose, {\em Quantum computation, entanglement and state reduction} , Phil. Trans. Roy. Soc. (London) A, 356,1927 (1998).




\bibitem{PowPer} W. L.  Power   and   I. C. Percival, {\em Decoherence of quantum wavepackets due to interaction with conformal spacetime fluctuations}, Proc.   Roy.   Soc.   Lond.   A456, 955 (2000).

\bibitem{AGB} L. Asprea, G. Gasbarri and A. Bassi, {\em Gravitational decoherence: a general non relativistic model}, arXiv:1905.01121.

\bibitem{GPP1} R. Gambini, R. Porto and J. Pullin, {\em Realistic Clocks, Universal Decoherence, and the Black Hole Information Paradox}, Phys. Rev. Lett. 93, 240401 (2004).

\bibitem{GPP2} R. Gambini, R. Porto and J. Pullin,{\em A relational solution to the problem of time in quantum mechanics and quantum gravity: a fundamental mechanism for quantum decoherence}, J. Phys. 6, 45 (2004).

 

\bibitem{Milburn1}  G.J. Milburn, {\em Intrinsic decoherence in quantum mechanics}, Phys. Rev. A 44, 5401 (1991).


\bibitem{Milburn2} G. J. Milburn, {\em Lorentz invariant intrinsic decoherence}, New J. Phys. 8, 96 (2006).

\bibitem{Boni} R. Bonifacio, {\em 	Time as a Statistical Variable and Intrinsic Decoherence}, Nuovo Cim. B114, 473-488 (1999).

\bibitem{AnHu08} C. Anastopoulos and B. L. Hu,{\em 
Intrinsic and Fundamental Decoherence: Issues and Problems }, Class. Quant. Grav. 25, 154003(2008).

\bibitem{AnHu07}C. Anastopoulos and B. L. Hu,
{\em Decoherence in quantum gravity: issues and critiques}, J. Phys.: Conf. Ser.67 012012 (2007).


\bibitem{XuBle20} Q. Xu and M.Blencowe, {\em Toy models for gravitational and scalar QED decoherence}, arXiv:2005.02554.

\bibitem{DSQL} M. Mohageg et al, {\em Deep Space Quantum Link} in Deep Space Gateway Science Workshop 2018, https://www.hou.usra.edu/meetings/deepspace2018/pdf/3039.pdf

\bibitem{event1} T.C.Ralph, G.J.Milburn and T.Downes, {\em Quantum connectivity of space-time and gravitationally induced decorrelation of entanglement},  Phys. Rev. A79,022121 (2009).

\bibitem{event2}T. J. Ralph and J. Pienar, {\em Quantum connectivity of space-time and gravitationally induced decorrelation of entanglement},  New J. Phys. 16, 085008 (2014).

 \bibitem{deutsch} D. Deutsch, {\em Quantum mechanics near closed timelike lines}, Phys.Rev.D.44, 3197 (1991)
 
 \bibitem{HawkingCP} S. W. Hawking, {\em Chronology protection conjecture}, Phys. Rev. D 46, 603 (1992).

\bibitem{KRW} B. S. Kay,  M. J. Radzikowski and R. M.  Wald,  {\em Quantum Field Theory on Spacetimes with a Compactly Generated Cauchy Horizon }, Comm. Math. Phys. 183, 533 (1997).

\bibitem{Wheeler} J. A. Wheeler,   {\em  Geons}, Phys. Rev.97, 511 (1955).


\bibitem{Davies}B. Davies, {\em Quantum Theory of Open Systems}, (Academic Press, London 1976).

\bibitem{BrePe07}H. P. Breuer and F. P. Petruccione, {\em The Theory of Open Quantum Systems} (Oxford University Press, 2007).


 \bibitem{OnWa1}T. Oniga and C.H.T. Wang, {\em Quantum gravitational decoherence of light and matter}, Phys. Rev. D93, 044027 (2016).
 

\bibitem{Ana96} C. Anastopoulos, {\em Quantum Theory of Non-Relativistic Particles Interacting with Gravity},  Phys. Rev.  D54, 1600   (1996).

\bibitem{Hab00}Z. Haba, {\em Decoherence by relic gravitons}, Mod. Phys. Lett. A15, 1519 (2000).

\bibitem{HaKl1} Z.Haba and H.Kleinert, {\em Quantum-Liouville and Langevin Equations for Gravitational Radiation Damping}, Int. J. Mod. Phys. A17, 3729 (2002).

\bibitem{RLLNJ} S. Reynaud, B. Lamine, A. Lambrecht. P. M. Neto and M.T. Jaekel, {\em Decoherence and gravitational backgrounds }, Int. J. Mod. Phys. A 17, 1003(2002). 

 

\bibitem{ReJa} S. Reynaud, B. Lamine, A. Lambrecht, P. Maia Neto, and M. T Jaekel, {\em HYPER and Gravitational Decoherence}, Gen.  Relativ.  Gravit. 362271, (2004).


\bibitem{Hu09} B. L. Hu, {\em Emergent/Quantum Gravity: Macro/Micro Structures of Spacetime}, J. Phys. Conf. Ser. 174, 012015 (2009).



\bibitem{FlHu}C. H. Fleming and  B. L. Hu, {\em Non-Markovian Dynamics of Open Quantum Systems: Stochastic Equations and their Perturbative Solutions}, Ann. Phys.  327, 1238 (2012).

\bibitem{FCAH} C. Fleming, N. I. Cummings, C. Anastopoulos and B. L. Hu, {\em The Rotating-Wave Approximation: Consistency and Applicability from an Open Quantum System Analysis },  J. Phys. A: Math. Theor. 43, 405304 (2010). 



\bibitem{Adler} S. Adler, {\em Quantum theory as an emergent phenomenon} (Cambridge University Press, Cambridge, 2004).

\bibitem{Diosi} L. Diosi, {\em Intrinsic time-uncertainties and decoherence: comparison of 4 models}, Braz. J. Phys. 35, 260  (2005).

\bibitem{Breuer} H. P. Breuer,   E.   G\"okl\"u,   and  C. L\"ammerzahl, {\em Metric fluctuations and decoherence}, Class.   Quantum   Grav. 26, 105012 (2009).



\bibitem{Thrn} K. S. Thorne, {\em Gravitational Waves},  arXiv:gr-qc/9506086.

\bibitem{QGP1} G. Amelino-Camelia, {\em Quantum-Spacetime Phenomenology},  Living Rev. Relativ. 16, 5 (2013).

\bibitem{QGP2} S. Hossenfelder, {\em Experimental Search for Quantum Gravity} (Springer, Berlin 2018)

\bibitem{Werner} R. Werner, {\em Screen observables in relativistic and nonrelativistic quantum mechanics}, J. Math. Phys. 27, 793 (1986).

\bibitem{Leon} J. Le\'on, {\em Time-of-arrival formalism for the relativistic particle}, J. Phys A: Math. Gen. 30, 4791 (1997).

\bibitem{AnSav12} C. Anastopoulos and N. Savvidou,  {\em Time-of-arrival Probabilities for General Particle Detectors}, Phys. Rev. A86, 012111 (2012).

\bibitem{AnSav19} C. Anastopoulos and N. Savvidou, {\em Time of arrival and localization of relativistic particles}, J. Math. Phys. 60, 032301 (2019).


\bibitem{Kief} C. Kiefer, {\em Decoherence in quantum electrodynamics and quantum gravity},    Phys. Rev. D46, 1658 (1992).


\bibitem{PWZ} M. Parikh, F. Wilczek, and G. Zahariade, {\em The Noise of Gravitons},  	Int. J. Mod. Phys. D29, 2042001  (2020).


\end{thebibliography}
\end{document}